\documentclass[graybox]{svmult}

\usepackage{type1cm}        
\usepackage{makeidx}         
\usepackage{graphicx}        
\usepackage{multicol}        
\usepackage[bottom]{footmisc}

\usepackage{txfonts}

\newcommand{\fig}[1]{Fig.~\ref{#1}}
\newcommand{\eq}[1]{Eq.~(\ref{#1})}
\newcommand{\eqs}[2]{Eqs.~(\ref{#1})-(\ref{#2})}

\makeindex             

\begin{document}

\title*{Vorticity and Spin Polarization in Heavy Ion Collisions: Transport Models}
\author{Xu-Guang Huang, Jinfeng Liao, Qun Wang, Xiao-Liang Xia}
\institute{Xu-Guang Huang \at  Physics Department, Center for Particle Physics and Field Theory, and Key Laboratory of Nuclear Physics and Ion-beam Application (MOE), Fudan University, Shanghai 200433, China, \email{huangxuguang@fudan.edu.cn}
\and Jinfeng Liao \at  Physics Department and Center for Exploration of Energy and Matter,
Indiana University, 2401 N Milo B. Sampson Lane, Bloomington, IN 47408, USA, \email{liaoji@indiana.edu}
\and Qun Wang \at  Department of Modern Physics, University of Science and Technology of China,
Hefei, Anhui 230026, China, \email{qunwang@ustc.edu.cn}
\and Xiao-Liang Xia \at Physics Department and Center for Particle Physics and Field Theory, Fudan University, Shanghai 200433, China, \email{xiaxl@fudan.edu.cn}}

%
%
\maketitle

\abstract{Heavy ion collisions generate strong fluid vorticty in the produced hot quark-gluon matter which could in turn induce measurable spin polarization of hadrons. We review recent progress on the vorticity formation and spin polarization in heavy ion collisions with transport models. We present an introduction to the fluid vorticity in non-relativistic and relativistic hydrodynamics and address various properties of the vorticity formed in heavy ion collisions. We discuss the spin polarization in a vortical fluid using the Wigner function formalism in which we derive the freeze-out formula for the spin polarization. Finally we give a brief overview of recent theoretical results for both the global and local spin polarization of $\Lambda$ and $\bar\Lambda$ hyperons.}

\section{Introduction}
\label{sec:intro}

Huge orbital angular momenta (OAM) are produced perpendicular to the reaction plane
in non-central high energy heavy ion collisions, and part of such huge OAM are transferred
to the hot and dense matter created in collisions \cite{Liang:2004ph,Liang:2004xn,Voloshin:2004ha,Betz:2007kg,Becattini:2007sr,Gao:2007bc,Huang:2011ru}. Due to the shear of the longitudinal flow particles with spins can be polarized via the spin-orbit coupling
in particle scatterings \cite{Liang:2004ph,Gao:2007bc,Huang:2011ru,Chen:2008wh}. Such a type of spin polarization with respect to
the reaction plane defined in the global laboratory frame of the collision is called the global polarization and is different from a particle's possible polarization
with respect to its production plane which depends on the particle's momentum \cite{Liang:2004ph}.
The global spin polarization of $\Lambda$ and $\bar{\Lambda}$ has been
measured by the STAR collaboration in Au+Au collisions over a wide range of beam energies, $\sqrt{s_{NN}}=7.7-200$
GeV \cite{STAR:2017ckg,Adam:2018ivw} and by ALICE collaboration in Pb+Pb collisions at 2.76 TeV and 5.02 TeV \cite{Acharya:2019ryw}. The magnitude of the global spin polarization
is about 2\% at 7.7 GeV which decreases to be about 0.3\% at 200 GeV and almost vanishes at LHC energies.

It has been shown that the spin-orbit coupling in microscopic particle scatterings
can lead to the spin-vorticity coupling in a fluid when taking an ensemble average over random
incoming momenta of colliding particles in a locally thermalized fluid \cite{Zhang:2019xya}.
In this way, the spin polarization is linked with the vorticity field in a fluid.
To describe the STAR data on the global polarization of hyperons, the hydrodynamic and
transport models have been used to calculate the vorticity field \cite{Baznat:2013zx,Csernai:2013bqa,Csernai:2014ywa,Becattini:2015ska,Teryaev:2015gxa,Jiang:2016woz,Deng:2016gyh,Ivanov:2017dff,Deng:2020ygd,Li:2017slc,Wei:2018zfb,Shi:2017wpk}.
In hydrodynamic models, the velocity and in turn the vorticity
fields in the fluid can be obtained naturally. In transport
models the phase space evolution of a multi-particle system is described by the Boltzmann transport equation with particle collisions, where  the position and momentum of each particle in the system at any time are explicitly known.
To extract the fluid velocity at one space-time point out of randomly distributed momenta in all events,
a suitable coarse-graining method has to be used that can map the transport description into hydrodynamic information~\cite{Jiang:2016woz,Deng:2016gyh}. The vorticity field can then be computed based on the so-obtained fluid velocity. Once the vorticity field is obtained, the
global polarization of hyperons can be calculated from an integral
over the freeze-out hyper-surface, which will be discussed in detail in Sec.~\ref{sec:pola} and Sec.~\ref{sec:pola:num}. The calculations following the above procedure give results on the global polarization that agree with the data \cite{Li:2017slc,Wei:2018zfb,Shi:2017wpk,Karpenko:2016jyx,Xie:2017upb,Sun:2017xhx,Xie:2016fjj}.

In this note we will give a brief review of vorticity formation and spin polarization in heavy ion collisions with transport models. We use the Minkowskian metric $g_{\mu\nu}={\rm diag}(1,-1,-1,-1)$ and natural unit $k_B=c=\hbar=1$ except for Sec.~\ref{sec:pola} in which $\hbar$ is kept explicitly.

\section{Fluid vorticity}
\label{sec:vort}

\subsection{Non-relativistic case}
\label{subsec:nonrel}
In non-relativistic hydrodynamics, the (kinematic) vorticity is a (pseudo)vector field that describes the local angular velocity of a fluid cell. Mathematically, it is defined as
\begin{eqnarray}
\label{defvor1}
\vec\omega({\vec x},t)=\frac{1}{2}\vec\nabla\times\vec v ({\vec x},t),
\end{eqnarray}
where $\vec v$ is the flow velocity with its three components denoted as $v_i$ ($i=1,2,3$). Sometimes it is also defined  without the pre-factor $1/2$ in Eq. (\ref{defvor1}). It can also be written in the tensorial form, $\omega_{ij}=(1/2)(\partial_i v_j-\partial_j v_i)$, so that $\omega_i=(1/2)\epsilon_{ijk}\omega_{jk}$, where $\epsilon_{ijk}$ is the three-dimension anti-symmetric tensor. For an ideal fluid, the flow is governed by the Euler equation which can be written in terms of $\vec\omega$ as,
\begin{eqnarray}
\label{eq:vort}
\frac{\partial\vec \omega}{\partial t}=\vec\nabla\times(\vec v\times\vec\omega).
\end{eqnarray}
This is called the vorticity equation. To arrive at \eq{eq:vort}, we have implicitly assumed the barotropic condition, $\vec\nabla\rho \mathbin{\parallel} \vec\nabla P$, which is satisfied if the pressure $P$ is a function of mass density $\rho$, $P=P(\rho)$. Equation (\ref{eq:vort}) has interesting consequences. Let us define the circulation integral of the velocity field over a loop $l$ co-moving with the fluid,
\begin{equation}
\label{eq:circ}
\Gamma=\oint_l\vec v\cdot d\vec x = 2\int_\Sigma\vec\omega \cdot d\vec\sigma ,
\label{circulation}
\end{equation}
where $\Sigma$ is a surface bounded by $l$ with $d\vec\sigma $ being its infinitesimal area element. Note that the second equality in \eq{circulation} follows from the Stokes theorem. It can be shown from \eq{eq:vort},
\begin{equation}
\frac{d\Gamma}{d\tau}=0,
\end{equation}
with co-moving time derivative $d/d\tau$. This result is called Helmholtz-Kelvin theorem which states that the vortex lines move with the fluid. Physically, it is equivalent to the angular momentum conservation for a closed fluid filament in the absence of viscosity, as all forces acting on the filament would be normal to it and generate no toque. Another interesting consequence of \eq{eq:vort} is the conservation of the flow helicity \cite{Moffatt:1969,Moreau:1961}
\begin{eqnarray}
\label{eq:heli}
{\cal H}_{\rm f}=\int d^3\vec x\,\vec\omega\cdot\vec v,
\end{eqnarray}
where the integral is over the whole space. Similar to energy, helicity is a quadratic invariant of the Euler equation of an ideal fluid although it is not positive definite.
In the following, we will generalize the notion of the vorticity to relativistic fluids and introduce the relativistic counterpart of the Helmholtz-Kelvin theorem and helicity conservation.

\subsection{Relativistic case}
\label{subsec:rel}
The generalization of vorticity to the relativistic case is not unique, and different definitions can be introduced for  different purposes. Here we discuss four types of relativistic vorticity. The first one is called the {\it kinematic vorticity} defined as
\begin{eqnarray}
\label{def:kv}
\omega_{\rm K}^\mu=\frac{1}{2}\epsilon^{\mu\nu\rho\sigma}u_\nu \partial_\rho u_\sigma,
\end{eqnarray}
which is a natural generalization of \eq{defvor1} as its spatial components recover \eq{defvor1} at non-relativistic limit. In the above, the four-velocity vector is defined by $u^\mu=\gamma(1,\vec v)$ with $\gamma=1/\sqrt{1-\vec v^2}$ the Lorentz factor. It is more convenient to define the kinematic vorticity tensor,
\begin{eqnarray}
\label{def:kvt}
\omega^{\rm K}_{\mu\nu}=-\frac{1}{2}(\partial_\mu u_\nu-\partial_\nu u_\mu),
\end{eqnarray}
so the kinematic vorticity vector is given by
\begin{equation}
\omega_{\rm K}^\mu=-(1/2)\epsilon^{\mu\nu\rho\sigma}u_\nu\omega^{\rm K}_{\rho\sigma}.
\label{vort-vec}
\end{equation}
Note that the minus sign in \eq{def:kvt} and (\ref{vort-vec}) is just a convention. The vorticity tensor and vector can also be defined without it. However, in either case (with or without the minus sign) the definition in  \eq{def:kv} always holds. We note that the relationship between the vorticity tensor and vector in \eq{vort-vec} also holds for the other types of vorticity definitions to be discussed below.

The second one is the {\it temperature vorticity} defined as
\begin{eqnarray}
\label{def:tv}
\omega^{\rm T}_{\mu\nu}=-\frac{1}{2}[\partial_\mu (T u_\nu)-\partial_\nu (T u_\mu)],
\end{eqnarray}
where $T$ is the temperature. The temperature vorticity for ideal neutral fluids is relevant to the relativistic version of Helmholtz-Kelvin theorem and helicity conservation~\cite{Becattini:2015ska,Deng:2016gyh}.
For an ideal neutral fluid, we can rewrite the Euler equation as
\begin{eqnarray}
\label{eq:euler}
(\varepsilon+P)\frac{d}{d\tau}u^\mu=\nabla^\mu P,
\end{eqnarray}
with $d/d\tau=u^\mu\partial_\mu$ and $\nabla_\mu=\partial_\mu-u_\mu (d/d\tau)$. The Euler equation (\ref{eq:euler}) can be put into the form of the Carter-Lichnerowicz equation with the help of the thermodynamic equation for a neutral fluid $dP=sdT$,
\begin{eqnarray}
\label{eq:cl}
\omega^{\rm T}_{\mu\nu}u^\nu=0,
\end{eqnarray}
from which the relativistic Helmholtz-Kelvin theorem can be obtained immediately,
\begin{eqnarray}
\label{eq:circ:r2}
\frac{d}{d\tau}\oint T u_\mu dx^\mu =2\oint \omega^{\rm T}_{\mu\nu} u^\mu dx^\nu=0.
\end{eqnarray}
Using \eq{eq:cl} we can also show that the temperature vorticity vector (multiplied by $T$) is conserved,
\begin{eqnarray}
\label{eq:rheli}
\partial_\mu (T\omega _{\rm T}^\mu ) = 4 u^\mu\omega_{\mu\nu}^{\rm T} \omega ^\nu_{\rm T} = 0,
\end{eqnarray}
where $\omega ^\mu_{\rm T}=-(1/2)\epsilon^{\mu\nu\rho\sigma}u_\nu\omega_{\rho\sigma}^{\rm T}$. The conserved charge ${\cal H}_{\rm T}=(1/2)\int d^3\vec x T^2 \gamma^2\vec v\cdot\vec\nabla\times\vec v$ is an extension of the helicity (\ref{eq:heli}) to the relativistic case for an ideal neutral fluid.

The third type is the charged-fluid counterpart of the temperature vorticity which we call the {\it enthalpy vorticity},
\begin{eqnarray}
\label{def:ev}
\omega^{\rm w}_{\mu\nu}=-\frac{1}{2}[\partial_\mu (w u_\nu)-\partial_\nu (w u_\mu)],
\end{eqnarray}
where $w=(\varepsilon+P)/n$ is the enthalpy per particle and $n$ is the charge density. In this case, the Euler equation (\ref{eq:euler}) can be written in the following Carter-Lichnerowicz form
\begin{eqnarray}
\label{eq:cl2}
u^\mu\omega^{\rm w}_{\mu\nu}=\frac{1}{2}T\nabla_\nu(s/n).
\end{eqnarray}
If the flow is isentropic ($s/n$ is a constant), we have $u^\mu\omega^{\rm w}_{\mu\nu}=0$, in the same form as \eq{eq:cl}. Therefore we have the conservation law for an ideal charged-fluid with the isentropic flow
similar to \eq{eq:circ:r2},
\begin{eqnarray}
\label{eq:circulation-w}
\frac{d}{d\tau}\oint w u_\mu dx^\mu =2\oint \omega^{\rm w}_{\mu\nu} u^\mu dx^\nu=0.
\end{eqnarray}
At the same time, the current $w\omega_{\rm w}^\mu$ is conserved, $\partial_\mu (w\omega_{\rm w}^\mu)=0$, and the corresponding conserved charge is the enthalpy helicity, ${\cal H}_{\rm w}=(1/2)\int d^3\vec x w^2 \gamma^2\vec v\cdot\vec\nabla\times\vec v$~\cite{Deng:2016gyh}.

The fourth vorticity is the {\it thermal vorticity}. It is defined as~\cite{Becattini:2015ska}
\begin{eqnarray}
\label{def:thv}
\omega^{\rm \beta}_{\mu\nu}=-\frac{1}{2}[\partial_\mu (\beta u_\nu)-\partial_\nu (\beta u_\mu)].
\end{eqnarray}
The thermal vorticity has an important property: for a fluid at global equilibrium, the four vector $\beta_\mu=\beta u_\mu$ is a Killing vector and is given by $\beta_\mu=b_\mu +\omega^{\rm \beta}_{\mu\nu} x^\nu$ with $b_\mu$ and $\omega^{\rm \beta}_{\mu\nu}$ constant. Thus, the thermal vorticity characterizes the global equilibrium of the fluid. In addition, the thermal vorticity is responsible for the local spin polarization of particles in a fluid at global equilibrium which we will discuss in details in the next section.

\section{Spin polarization in a vortical fluid}
\label{sec:pola}
A semi-classical way to describe the space-time evolution of spin
degrees of freedom is through the spin-dependent distribution function.
The quantum theory provides a more rigorous description for the spin
evolution through the Wigner function, a quantum counterpart of the
distribution function. For a relativistic spin-1/2 fermion, one has
to use the covariant Wigner function \cite{Heinz:1983nx,Elze:1986qd,Vasak:1987um,Zhuang:1995pd},
which is a $4\times4$ matrix function of position and momentum. Now
the covariant Wigner function becomes a useful tool to study the chiral
magnetic and vortical effect and other related effects
\cite{Gao:2012ix,Chen:2012ca,Gao:2015zka,Fang:2016vpj,Hidaka:2016yjf,Mueller:2017lzw,Huang:2018wdl,Liu:2018xip}.
The Wigner function is equivalent to the quantum field and contains
all information that the quantum field does. Therefore the spin information
in phase space is fully encoded in the Wigner function from which
one can obtain the quark polarization from its axial vector components.

The covariant Wigner function for spin-1/2 fermions in an external
electromagnetic field is defined by \cite{Heinz:1983nx,Elze:1986qd,Vasak:1987um,Zhuang:1995pd}
\begin{equation}
W_{\alpha\beta}(x,p)=\frac{1}{(2\pi)^{4}}\int d^{4}ye^{-ip\cdot y}\left\langle \bar{\psi}_{\beta}\left(x+\frac{y}{2}\right)U\left(A;x+\frac{1}{2}y,x-\frac{1}{2}y\right)\psi_{\alpha}\left(x-\frac{y}{2}\right)\right\rangle ,
\end{equation}
where $\psi_{\alpha}$ and $\bar{\psi}_{\beta}$ are the fermionic
field components ($\alpha,\beta=1,2,3,4$ are the spinor indices),
$U(A;x_{2},x_{1})=\exp\left[iQ\int_{x_{1}}^{x_{2}}dx^{\mu}A_{\mu}(x)\right]$
is the gauge link that makes gauge invariance of the Wigner function
with $A_{\mu}$ being the electromagnetic gauge potential, and $\left\langle \hat{O}\right\rangle $
denotes the ensemble average of the operator $\hat{O}$ over thermal
states. As a $4\times4$ complex matrix having 32 real variables,
the Wigner function satisfies $W^{\dagger}=\gamma_{0}W\gamma_{0}$,
which reduces the number of independent variables to 16.
Therefore the Wigner function can be expanded in terms of 16 generators of Clifford
algebra $\{1,\gamma_{5},\gamma^{\mu},\gamma_{5}\gamma^{\mu},\sigma^{\mu\nu}\}$
with $\gamma^{5}\equiv i\gamma^{0}\gamma^{1}\gamma^{2}\gamma^{3}$
and $\sigma^{\mu\nu}\equiv\frac{i}{2}[\gamma^{\mu},\gamma^{\nu}]$,
\begin{equation}
W=\frac{1}{4}\left(\mathcal{F}+i\gamma^{5}\mathcal{P}+\gamma^{\mu}\mathcal{V}_{\mu}+\gamma^{5}\gamma^{\mu}\mathcal{A}_{\mu}+\frac{1}{2}\sigma^{\mu\nu}\mathcal{S}_{\mu\nu}\right),
\end{equation}
where the coefficients are the scalar ($\mathcal{F}$), pseudoscalar
($\mathcal{P}$), vector ($\mathcal{V}_{\mu}$), axial vector ($\mathcal{A}_{\mu}$)
and tensor ($\mathcal{S}_{\mu\nu}$) components with 1, 1, 4, 4 and
6 independent variables, respectively. Each component of $W$ can be
extracted by multiplying it with the corresponding generator and taking
a trace. These components are all real functions of phase space coordinates
and satisfy 32 real equations with 16 redundant equations.
For massless fermions, the equations for the vector and axial-vector component
are decoupled from the rest components. They can be linearly combined into
the right-handed and left-handed vector component, both sectors satisfy
the same set of equations. By solving the set of equations one can derive
the right-handed and left-handed currents which give the chiral magnetic and vortical effect
in an external electromagnetic field and a vorticity field
\cite{Gao:2012ix,Chen:2012ca,Gao:2015zka,Hidaka:2016yjf,Huang:2018wdl,Liu:2018xip,Gao:2017gfq,Gao:2018wmr}.
For massive fermions, the equations for Wigner function components
are all entangled and hard to solve. Fortunately there is a natural expansion parameter
in these equations, the Planck constant $\hbar$, which gives the
order of quantum correction. The Wigner function components can thus
be obtained by solving these questions order by order in $\hbar$,
which is called semi-classical expansion
\cite{Weickgenannt:2019dks,Gao:2019znl,Hattori:2019ahi,Wang:2019moi,Liu:2020flb,Sheng:2020oqs,Florkowski:2018ahw,Yang:2020hri,Weickgenannt:2020aaf,Wang:2020pej}.

The Wigner function components at the zero-th order in $\hbar$ are
given by \cite{Weickgenannt:2019dks}
\begin{eqnarray}
\mathcal{F}^{(0)}(x,p) & = & m\delta(p^{2}-m^{2})V^{(0)}(x,p),\nonumber \\
\mathcal{P}^{(0)}(x,p) & = & 0,\nonumber \\
\mathcal{V}_{\mu}^{(0)}(x,p) & = & p_{\mu}\delta(p^{2}-m^{2})V^{(0)}(x,p),\nonumber \\
\mathcal{A}_{\mu}^{(0)}(x,p) & = & mn_{\mu}^{(0)}(x,\mathbf{p})\delta(p^{2}-m^{2})A^{(0)}(x,p),\nonumber \\
\mathcal{S}_{\mu\nu}^{(0)}(x,p) & = & m\Sigma_{\mu\nu}^{(0)}(x,\mathbf{p})\delta(p^{2}-m^{2})A^{(0)}(x,p),\label{eq:zero-th-comp}
\end{eqnarray}
with
\begin{eqnarray}
V^{(0)}(x,p) & \equiv & \frac{2}{(2\pi\hbar)^{3}}\sum_{e,s=\pm}\theta(ep^{0})f_{s}^{(0)e}(x,e\mathbf{p}),\nonumber \\
A^{(0)}(x,p) & \equiv & \frac{2}{(2\pi\hbar)^{3}}\sum_{e,s=\pm}s\theta(ep^{0})f_{s}^{(0)e}(x,e\mathbf{p}),\nonumber \\
n^{(0)\mu}(x,p) & \equiv & \theta(p^{0})n^{+\mu}(x,\mathbf{p})-\theta(-p^{0})n^{-\mu}(x,\mathbf{p}),\nonumber \\
\Sigma_{\mu\nu}^{(0)}(x,p) & = & -\frac{1}{m}\epsilon_{\mu\nu\alpha\beta}p^{\alpha}n^{(0)\beta},\label{eq:v0-a0-n0-sig0}
\end{eqnarray}
where $e=\pm$ denotes particle/antiparticle, $s=\pm$ denotes spin
up/down, $f_{s}^{(0)e}$ are the distribution functions. In Eq. (\ref{eq:v0-a0-n0-sig0})
$n^{\mu}(\mathbf{p},\mathbf{n})$ is the spin four-vector and $n^{\pm\mu}(x,\mathbf{p})$
are spin-four vector for particle/antiparticle given by
\begin{eqnarray}
n^{+\mu}(x,\mathbf{p}) & = & \left(\frac{\mathbf{n}^{+}\cdot\mathbf{p}}{m},\mathbf{n}^{+}+\frac{\mathbf{n}^{+}\cdot\mathbf{p}}{m(m+E_{\mathbf{p}})}\mathbf{p}\right),\nonumber \\
n^{-\mu}(x,\mathbf{p}) & = & \left(\frac{\mathbf{n}^{-}\cdot\mathbf{p}}{m},-\mathbf{n}^{-}-\frac{\mathbf{n}^{-}\cdot\mathbf{p}}{m(m+E_{\mathbf{p}})}\mathbf{p}\right),\label{nDef}
\end{eqnarray}
where $\mathbf{n}^{\pm}$ are spin quantization directions for particle/antiparticle
in the particle's rest frame. In general $\mathbf{n}^{+}$ can be
different from $\mathbf{n}^{-}$. We note that $n^{+\mu}(x,\mathbf{p})$
can be expressed by a Lorentz boost from the the particle's rest frame
to the lab frame in which the particle has the momentum $\mathbf{p}$
\begin{eqnarray}
n^{+\mu}(x,\mathbf{p}) & = & \Lambda_{\;\nu}^{\mu}(-\mathbf{v}_{p})n^{+\nu}(\mathbf{0},\mathbf{n}^{+}).\label{eq:n-part}
\end{eqnarray}
Here $\Lambda_{\;\nu}^{\mu}(-\mathbf{v}_{p})$ is the Lorentz transformation
for $\mathbf{v}_{p}=\mathbf{p}/E_{p}$ and $n^{+\nu}(\mathbf{0},\mathbf{n}^{+})=(0,\mathbf{n}^{+})$
is the four-vector of the spin quantization direction in the particle's
rest frame. One can check that $n^{+\mu}(x,\mathbf{p})$ satisfies
$n_{\mu}^{+}n_{+}^{\mu}=-1$ and $n^{+}\cdot p=0$. Similarly $n^{-\mu}(x,\mathbf{p})$
for the antiparticle can be expressed by
\begin{eqnarray}
n^{-\mu}(x,\mathbf{p}) & = & \Lambda_{\;\nu}^{\mu}(\mathbf{v}_{p})n^{-\nu}(\mathbf{0},\mathbf{n}^{-}),\label{eq:n-antipart}
\end{eqnarray}
where $n^{-\nu}(\mathbf{0},\mathbf{n}^{-})=(0,-\mathbf{n}^{-})$.

We see in Eqs. (\ref{eq:zero-th-comp},\ref{eq:v0-a0-n0-sig0}) that
the axial vector component corresponds to the spin four-vector. We
can rewrite the last line of Eq. (\ref{eq:v0-a0-n0-sig0}) in another
form \cite{Weickgenannt:2019dks}
\begin{equation}
n_{\mu}^{(0)}=-\frac{1}{2m}\epsilon_{\mu\nu\alpha\beta}p^{\nu}\Sigma^{(0)\alpha\beta},
\end{equation}
where $n_{\mu}^{(0)}$ is the Pauli-Lubanski pseudovector and $\Sigma^{(0)\alpha\beta}$
plays the role of a spin angular momentum tensor.

At the first order in $\hbar$, the axial vector component is \cite{Fang:2016vpj,Weickgenannt:2019dks}
\begin{equation}
\mathcal{A}_{\mu}^{(1)}=m\bar{n}_{\mu}^{(1)}\delta(p^{2}-m^{2})+\tilde{F}_{\mu\nu}p^{\nu}V^{(0)}\delta^{\prime}(p^{2}-m^{2}),\label{eq:1st-axial-vector}
\end{equation}
where $\tilde{F}_{\mu\nu}=(1/2)\epsilon_{\mu\nu\alpha\beta}F^{\alpha\beta}$
and
\begin{equation}
\bar{n}_{\mu}^{(1)}\equiv-\frac{1}{2m}\epsilon_{\mu\nu\alpha\beta}p^{\nu}\bar{\Sigma}^{(1)\alpha\beta},\label{eq:n1-sig-1}
\end{equation}
is the first-order on-shell correction to $n_{\mu}^{(0)}A^{(0)}$.
In Eq. (\ref{eq:n1-sig-1}) $\bar{\Sigma}^{(1)\alpha\beta}$ can be
decomposed as
\begin{equation}
\bar{\Sigma}^{(1)\alpha\beta}=\frac{1}{2}\chi^{\alpha\beta}+\Xi^{\alpha\beta},
\end{equation}
where the tensor $\Xi^{\alpha\beta}$ is symmetric and satisfies $p_{\alpha}\Xi^{\alpha\beta}=0$.
The evolution equations for $\chi^{\alpha\beta}$ and for $\Xi^{\alpha\beta}$
are \cite{Weickgenannt:2019dks}
\begin{eqnarray}
p\cdot\nabla^{(0)}\chi_{\mu\nu} & = & 0,\nonumber \\
p\cdot\nabla^{(0)}\Xi_{\mu\nu} & = & F_{\ \mu}^{\alpha}\Xi_{\nu\alpha}-F_{\ \nu}^{\alpha}\Xi_{\mu\alpha},
\end{eqnarray}
where $\nabla^{(0)\mu}\equiv\partial_{x}^{\mu}-F^{\mu\nu}\partial_{p\nu}$.
The component $\chi_{\mu\nu}$ satisfies the constraint
\begin{equation}
p^{\nu}\chi_{\mu\nu}=\nabla_{\mu}^{(0)}V^{(0)}.
\end{equation}
In global equilibrium a special choice of $\chi_{\mu\nu}$ is
\begin{equation}
\chi_{\mu\nu}=-\omega_{\mu\nu}^{\beta}\frac{\partial V^{(0)}}{\partial(\beta p_{0})},
\end{equation}
where $\omega_{\mu\nu}^{\beta}$ is the thermal vorticity tensor (\ref{def:thv}) and
\begin{eqnarray}
V^{(0)} & \equiv & \frac{2}{(2\pi\hbar)^{3}}\sum_{s}\left[\theta(u\cdot p)f_{s}^{(0)+}+\theta(-u\cdot p)f_{s}^{(0)-}\right],\nonumber \\
f_{s}^{(0)\pm} & = & \frac{1}{\exp(\beta u\cdot p\mp\beta\mu_{s})+1}.\label{eq:omega}
\end{eqnarray}
Here $u^{\mu}$ is the flow velocity and $\omega_{\mu\nu}$ is the
vorticity tensor. Therefore the vorticity dependent part of the axial
vector component in Eq. (\ref{eq:1st-axial-vector}) reads \cite{Fang:2016vpj,Weickgenannt:2019dks}
\begin{equation}
\mathcal{A}_{\mu}^{(1)}=\frac{1}{4}\epsilon_{\mu\nu\rho\sigma}p^{\nu}\omega_{\beta}^{\rho\sigma}\frac{\partial V^{(0)}}{\partial(\beta u\cdot p)}\delta(p^{2}-m^{2}).
\end{equation}
We can integrate $\mathcal{A}_{\mu}^{(1)}$ over $p_{0}$ to make
the momentum of the particle/antiparticle to be on the mass shell.
The average spin per particle (with an additional factor 1/2 from
the particle's spin) is given by
\begin{equation}
S_{\mu}^{\pm}=-\frac{1}{8(u\cdot p)}\epsilon_{\mu\nu\rho\sigma}p^{\nu}\omega_{\beta}^{\rho\sigma}(1-f_{\mathrm{FD}}^{\pm}),\label{eq:spin-vector}
\end{equation}
where $f_{\mathrm{FD}}^{\pm}$ is the on-shell Fermi-Dirac distribution
function with $p_{0}$ replaced by $\pm E_{p}$ ($E_{p}\equiv\sqrt{m^{2}+\mathbf{p}^{2}}$)
in $f_{s}^{(0)\pm}$ for a particle/antiparticle respectively. We
can generalize the above equilibrium formula to a hydrodynamic process
at a freeze-out hypersurface $\sigma_{\mu}$ \cite{Fang:2016vpj,Liu:2020flb,Becattini:2013fla},
in this case the average spin per particle is given by
\begin{equation}
S^{\mu}(p)=-\frac{1}{8}\epsilon^{\mu\nu\rho\sigma}p_{\nu}\frac{\int d\sigma_{\lambda}p^{\lambda}\omega_{\rho\sigma}^{\beta}(u\cdot p)^{-1}f_{\mathrm{FD}}(1-f_{\mathrm{FD}})}{\int d\sigma_{\lambda}p^{\lambda}f_{\mathrm{FD}}},
\label{average-spin-vector}
\end{equation}
where we have suppressed the index $\pm$ for the particle/antiparticle
since the above formula is valid for both particles and antiparticles.
If the momentum is not large compared with the particle mass, we have
$u\cdot p\approx m$ and Eq. (\ref{average-spin-vector}) recovers
the result in Ref. \cite{Fang:2016vpj,Liu:2020flb,Becattini:2013fla} which is
widely used in calculating the hadron polarization in heavy ion collisions.

\section{Vorticity in heavy ion collisions}
\label{sec:vort:num}

There are multiple sources of vorticity in heavy ion collisions.
One source is the global orbital angular momentum (OAM) of the two colliding nuclei in non-central collisions. Geometrically, this OAM is perpendicular to the reaction plane~\footnote{Strictly speaking, this is true only after taking average over many collision events, as the collision geometry itself (and thus the direction of the OAM) suffers from event-by-event fluctuations.}. After the collision, a fraction of the total OAM is retained in the produced quark-gluon matter and induces vorticity. As we will discuss later in this section, in the mid-rapidity region for $\sqrt{s}$ larger than about 10 GeV, such a generated vorticity decreases with the increasing beam energy, consistent with the measured global spin polarization of $\Lambda$ and $\bar{\Lambda}$ hyperons. The second source of the vorticity is the jet-like fluctuation in the fireball which can induce smoke-loop type vortex around the fast moving particle~\cite{Betz:2007kg}. The direction of such vorticity is not correlated to the reaction plane and thus does not contribute to the global $\Lambda$ polarization. Instead, on an event-by-event basis, it generates a near-side longitudinal spin-spin correlation~\cite{Pang:2016igs}. The third source of the vorticity is the inhomogeneous expansion of the fire ball~\cite{Jiang:2016woz,Wei:2018zfb,Pang:2016igs,Becattini:2017gcx,Xia:2018tes}. In particular, anisotropic flows in the transverse plane can produce a quadrupole pattern of the longitudinal vorticity along the beam direction while the inhomogeneous transverse expansion can produce a transverse vorticity circling the longitudinal axis. There may be other sources of vorticity, e.g., the strong magnetic field created by fast-moving spectators may magnetize the quark-gluon matter and potentially  lead to a vorticity along the direction of the magnetic field through the so-called Einstein-de Haas effect.

Vorticity formation in high energy nuclear collisions has been extensively studied in relativistic hydrodynamic models, such as ECHO-QGP \cite{Becattini:2015ska}, PICR \cite{Csernai:2013bqa,Csernai:2014ywa} and CLVisc \cite{Pang:2016igs} in (3+1) dimensions. Using the ECHO-QGP code \cite{DelZanna:2013eua}, different vorticities in relativistic hydrodynamics are studied in the context of directed flow in non-central collisions \cite{Becattini:2015ska}. The evolution of the kinematic vorticity has been calculated using the PICR hydrodynamic code \cite{Csernai:2013bqa}. Using CLVisc \cite{Pang:2012he,Pang:2018zzo} with event-by-event fluctuating initial conditions the vorticity distributions have been calculated. A structure of vortex-pairing in the transverse plane due to the convective flow of hot spots in the radial direction is found to possibly form in high energy heavy ion collisions.

In this section we will focus on the kinematic and thermal vorticity based on transport models such as the AMPT model, but the discussion will also involve other types of vorticity. Before we go into the details, let us first discuss the setup of numerical simulations for extracting vorticity structures from the AMPT model~\cite{Jiang:2016woz} as well as the HIJING model~\cite{Deng:2016gyh} with partons as basic degrees of freedom.

\subsection{Setup of computation in transport models}
\label{sec:setup}

According to the definitions in Sec.~\ref{sec:vort}, in order to calculate the kinematic and thermal vorticity, we first need to obtain the velocity field $u^\mu$ (with normalization $u^\mu u_\mu=1$) and the temperature field $T$. A natural way to achieve this is by using the energy momentum tensor $T^{\mu\nu}$ through which we can define the velocity field and the energy density $\varepsilon$ as the eigenvector and eigenvalue of $T^{\mu\nu}$ respectively,
\begin{eqnarray}
\label{eq:T:u}
T^{\mu\nu} u_\nu=\varepsilon u^\mu .
\end{eqnarray}
The temperature $T$ can be determined from $\varepsilon$ as a function of $T$ by assuming a local equilibrium. In transport models such as HIJING, AMPT or UrQMD, the position and momentum of each particle  is known at any moment. A simple way to determine $T^{\mu\nu}$ as a function of space-time is by the coarse-grained method. This is done by splitting the whole space-time volume into grid cells and calculating an event average of $\sum _i p^\mu_i p^\nu_i/p^0_i$ inside each space-time cell,
\begin{equation}
T^{\mu\nu}(x) = \frac{1}{\Delta x \Delta y \Delta z}\left\langle
\sum _i \frac{p^\mu_i p^\nu_i}{p^0_i} \right\rangle
\end{equation}
where $i$ labels a particle inside the cell. The event average is taken to cancel the random or thermal motion of particles in each space-time cell, and finally the collective motion is kept.

Another way is to introduce a function $\Phi(x,x_i)$ to smear a physical quantity (such as the momentum) of the $i$-th particle at $x_i$ in an event. In such a way, we can construct a continuous function of that physical quantity~\cite{Deng:2016gyh,Wei:2018zfb}. Physically, function $\Phi(x, x_i)$ reflects the quantum nature of the particle as a wave-packet. With $\Phi(x, x_i)$, the phase space distribution can be obtained as
\begin{eqnarray}
\label{eq:dist}
f(x,p)=\frac{1}{\cal N}\sum_{i}(2\pi)^3\delta^{(3)}[\vec p-\vec p_i(t)]\Phi[x,x_i(t)],
\end{eqnarray}
where ${\cal N}=\int d^3\vec x \Phi(x,x_i)$ is a normalization factor. Then the energy-momentum tensor is given by
\begin{eqnarray}
\label{eq:EM}
T^{\mu\nu}(x)=\int\frac{d^3\vec p}{(2\pi)^3}\frac{p^\mu p^\nu}{p^0} f(x,p)=\frac{1}{\cal N}\sum_{i}\frac{p^\mu_i p^\nu_i}{p^0_i}\Phi(x,x_i).
\end{eqnarray}

The choice of the smearing function is important. Here we give two examples.

(a) The $\Delta$ smearing. This is given by generalizing the $\delta$ function $\delta^{(3)}[x-x_i(t)]$ (corresponding to a zero smearing) to
\begin{eqnarray}
\label{smear:Delt}
\Phi_\Delta[x,x_i(t)]=\delta_\Delta^{(3)}[\vec x-\vec x_i(t)],
\end{eqnarray}
which is $1$ if $|x-x_i(t)|<\Delta x, |y-y_i(t)|<\Delta y, |z-z_i(t)|<\Delta z$, and is $0$ otherwise.
This is actually the coarse-grained method as we have discussed ealier in this subsection.

\begin{figure}[!htb]
\begin{center}
\includegraphics[width=4.5cm]{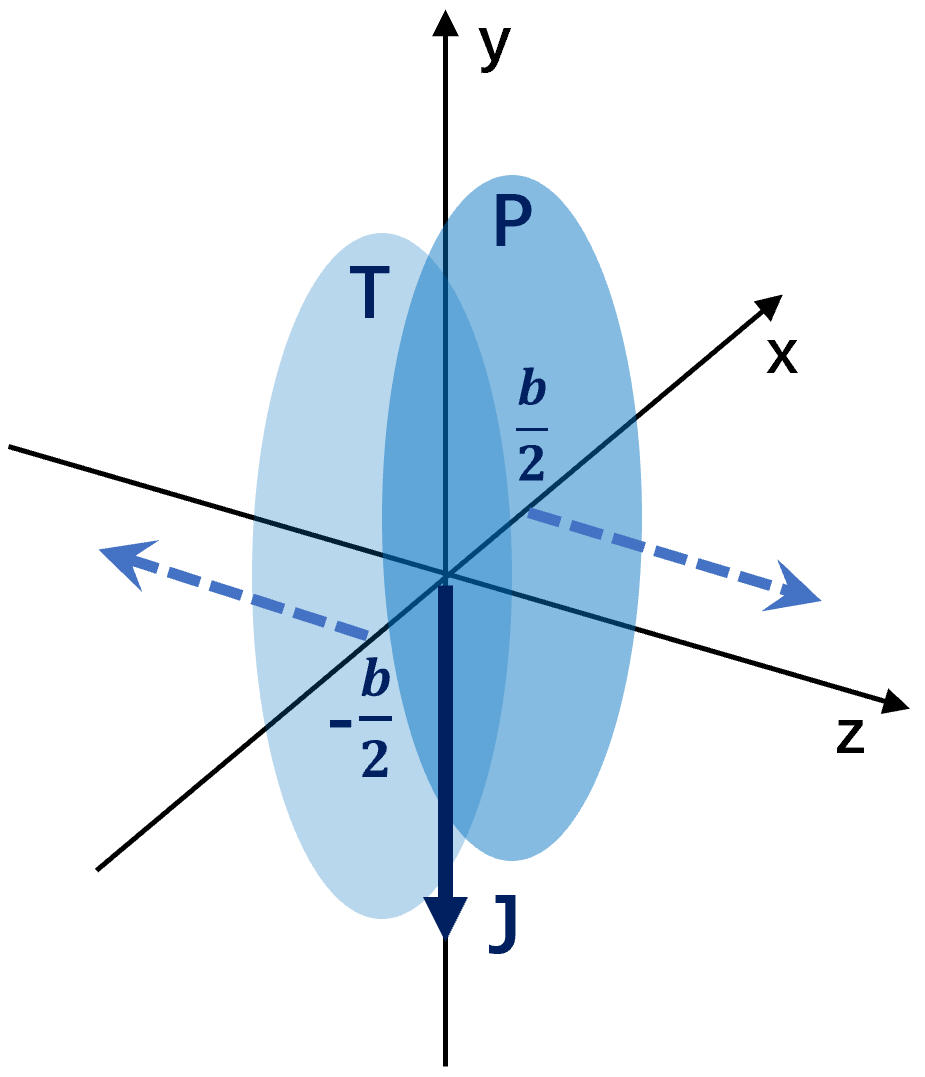}
\caption{The coordinate system of a heavy ion collision. Here, `T' is for target and `P' is for projectile.}
\label{frame}
\end{center}
\end{figure}

(b) The Gaussian smearing \cite{Deng:2016gyh,Pang:2012he,Hirano:2012kj}. This is given by
\begin{eqnarray}
\label{smear:Gaus}
\Phi_{\rm G}[x,x_i(\tau)]=K \exp\left[-\frac{(x-x_i)^2}{2\sigma_x^2}-\frac{(y-y_i)^2}{2\sigma_y^2}-\frac{(\eta-\eta_i)^2}{2\sigma_\eta^2}\right],
\end{eqnarray}
where we have adopted the Milne coordinate $(\tau, x, y, \eta)$ with $\eta=(1/2)\ln[(t+z)/(t-z)]$ being the spacetime rapidity and $\tau=\sqrt{t^2-z^2}$ being the proper time instead of the Minkowski coordinate, and
$K$ and $\sigma_{x,y,z}$ are parameters that can be determined by fitting to experimental data. As a  convention for the coordinate system, the $z$-axis is along the beam direction of the projectile, the $x$-axis is along the impact parameter from the target to the projectile nucleus, and the $y$-axis is along
$\hat{\vec z}\times\hat{\vec x}$, see \fig{frame}.

\subsection{Results for kinematic vorticity}
\label{sec:num:kv}
The kinematic vorticity (\ref{def:kv}) is a natural extension of the non-relativistic vorticity (\ref{defvor1}) which is a direct measure of the angular velocity of the fluid cell.  We will discuss a series of features of the kinematic vorticity (including the non-relativistic one).

\textit{Centrality dependence.}
\begin{figure}[!htb]
\begin{center}
\includegraphics[width=5.5cm]{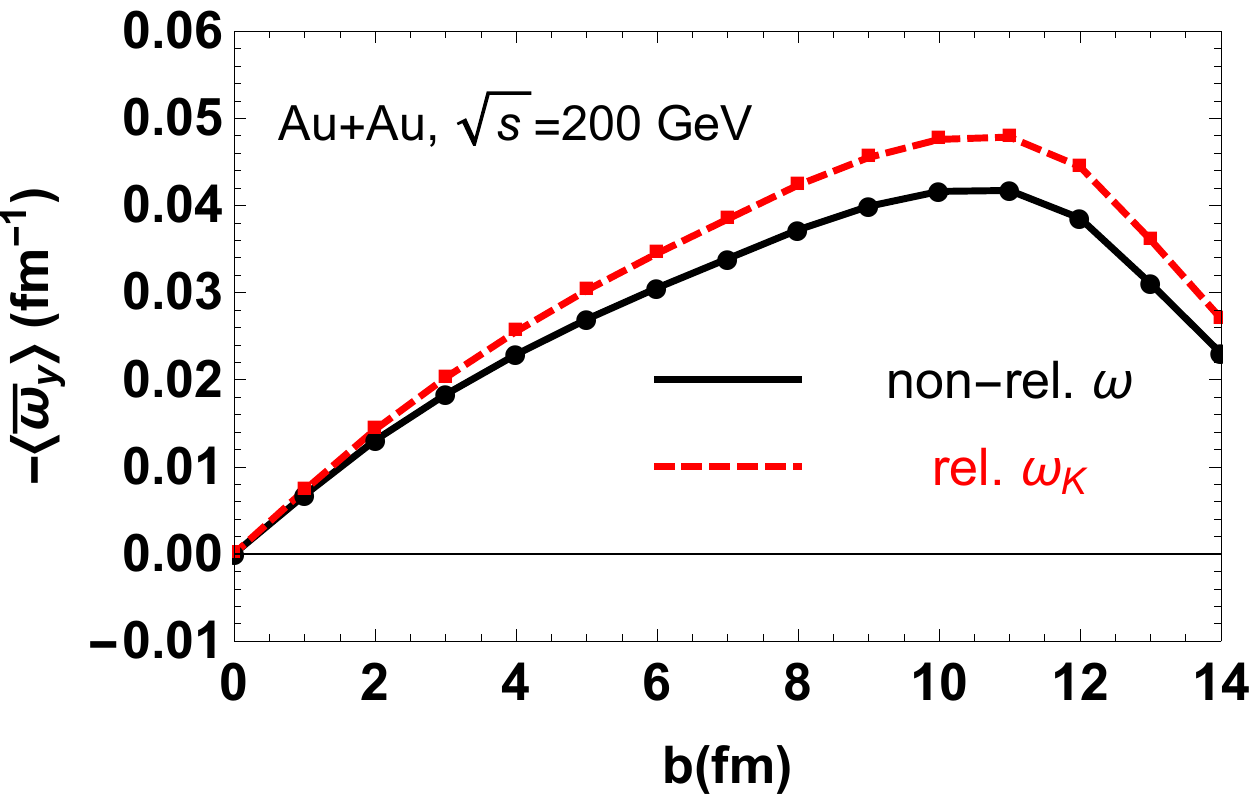}
\includegraphics[width=5.5cm]{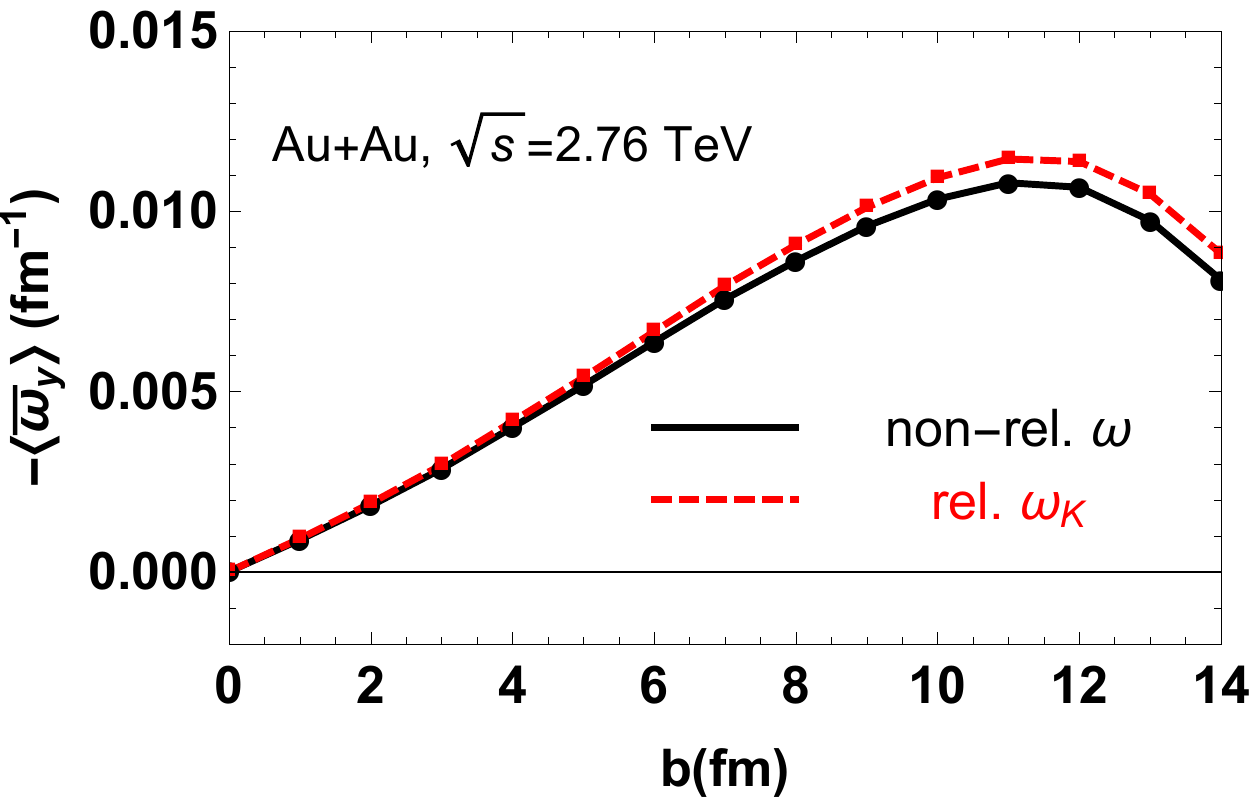}
\caption{The $y$-components of the non-relativistic vorticity in Eq. (\ref{defvor1}) and the relativistic kinematic vorticity in Eq. (\ref{def:kv}) at $\tau=\tau_0$ and $\eta=0$ in $200$ GeV Au + Au and $2.76$ TeV Pb + Pb collisions ~\cite{Deng:2016gyh}.}
\label{bdepen}
\end{center}
\end{figure}
It is expected that for a given collision energy, the total angular momentum of the two colliding nuclei with respect to the collision center increases with the centrality or equivalently impact parameter. As a result, the vorticity is expected to increase with the centrality too. This is indeed the case as shown in \fig{bdepen} in which the average non-relativistic and relativistic vorticity in $y$-direction $\langle\overline{\omega}_y\rangle$ at initial time ($\tau_0=0.4$ fm for $\sqrt{s}=200$ GeV and $\tau_0=0.2$ fm for $\sqrt{s}=2.76$ TeV) and mid-rapidity are plotted as functions of the impact parameter $b$. The average is over both the transverse overlapping region (indicated by an overline of $\omega_y$) and the collision events (indicated by $\langle\cdots\rangle$), see Ref. \cite{Deng:2016gyh} for details. We see that the magnitude of the kinematic vorticity is big, for example, $|\omega_y|$ is about $10^{20} s^{-1}$ at $b=10$ fm and $\sqrt{s}=200$ GeV, a value surpassing the vorticity of any other known fluids. We also notice that the kinematic vorticity begins to decrease with $b$ when $b\gtrsim 2R_A$ with $R_A$ being the nucleus radius, reflecting the fact that the two colliding nuclei begin to separate.

\begin{figure}[!htb]
\begin{center}
\includegraphics[width=6cm]{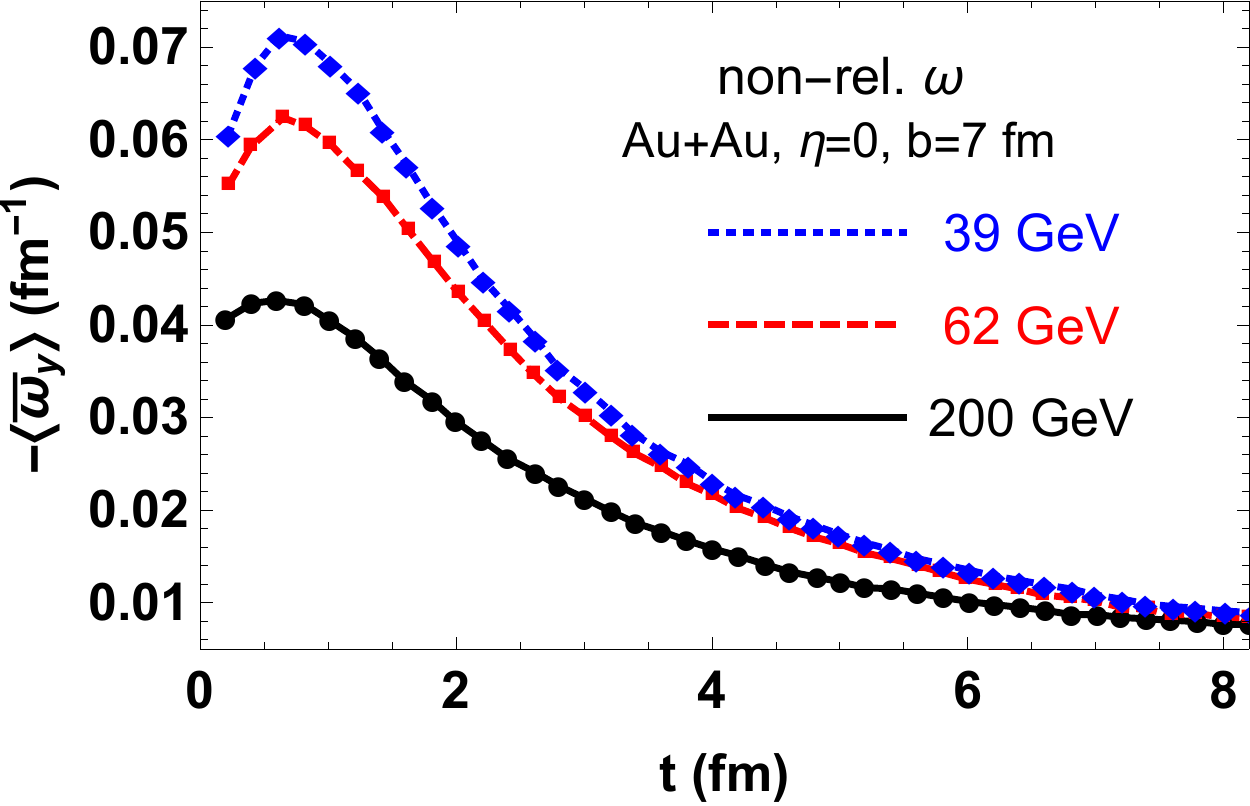}
\caption{The time evolution of the non-relativistic vorticity for different collision energies~\cite{Jiang:2016woz}.}
\label{tdepen}
\end{center}
\end{figure}

\textit{Energy dependence.}
\begin{figure}[!htb]
\begin{center}
\includegraphics[width=5.5cm]{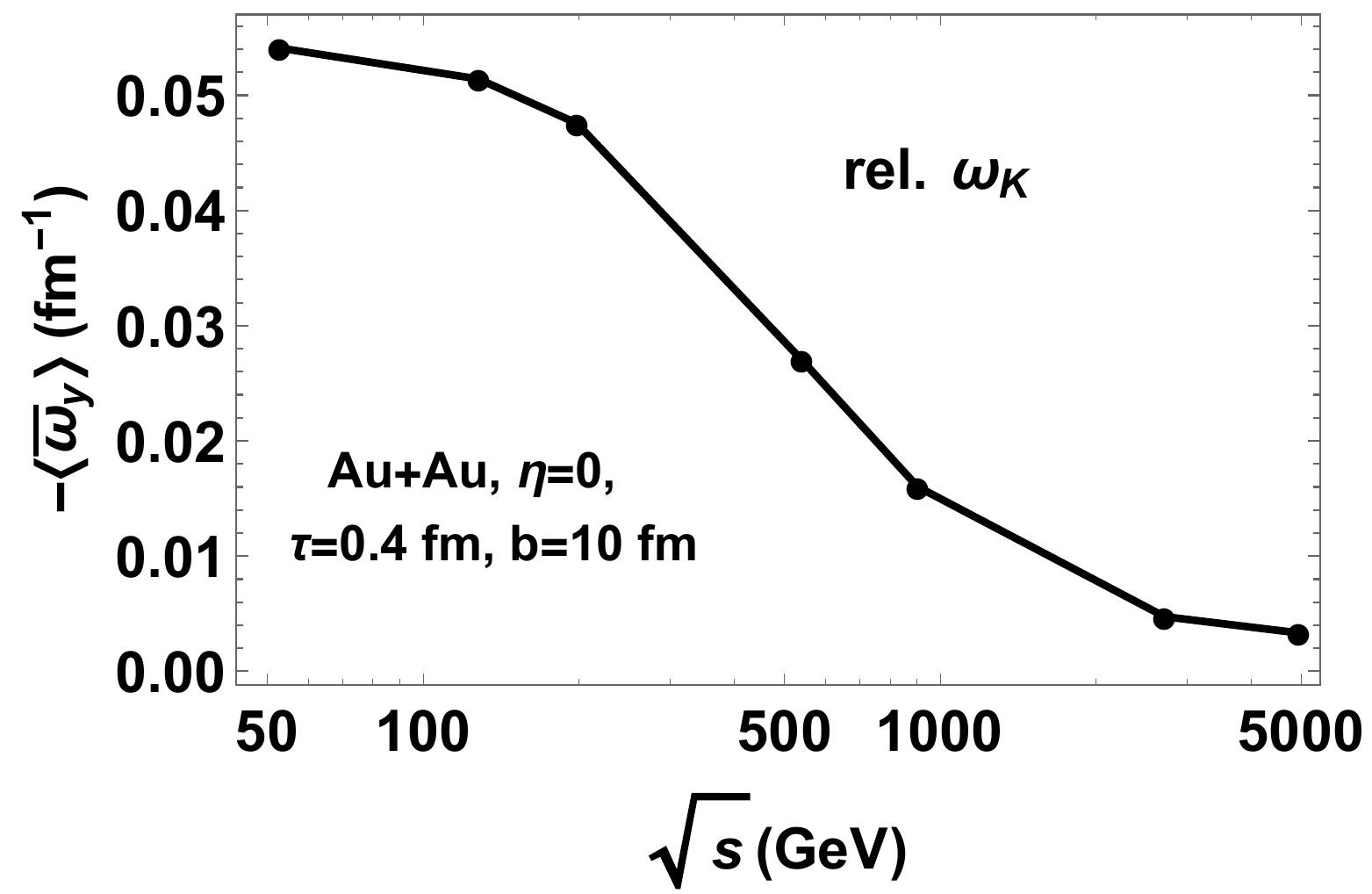}
\caption{The collision energy dependence of the kinematic vorticity at mid-rapidity~\cite{Deng:2016gyh}. }
\label{edepen}
\end{center}
\end{figure}
It is obvious that the total angular momentum of the two colliding nuclei grows with collision energy $\sqrt{s}$ at a fixed impact parameter. Naively one would then expect a similar energy dependence of the vorticity. However, as shown in \fig{edepen} (and also in \fig{tdepen}), the $y$-component of the kinematic vorticity at mid-rapidity decreases as $\sqrt{s}$ increases. Such a behavior features the relativistic effect in the mid-rapidity region: as $\sqrt{s}$ increases, two nuclei become more transparent to each other and leave the mid-rapidity region more boost invariant which supports a lower vorticity. To put it in another way: while the total angular momentum of the colliding system increases with the beam energy, the fraction of that angular momentum carried by the fireball at mid-rapidity decreases rapidly with the beam energy~\cite{Jiang:2016woz}. At low energy, the relativistic effect becomes less important and the fireball acquires a considerably more fraction of the system's angular momentum, leading to a much increased vorticity~\cite{Jiang:2016woz,Deng:2016gyh}. At very low energy, however, the total angular momentum would be small and the vorticity become inevitably small again \cite{Deng:2020ygd}.

\textit{Correlation to the participant plane.}
\begin{figure}[!htb]
\begin{center}
\includegraphics[width=5.5cm]{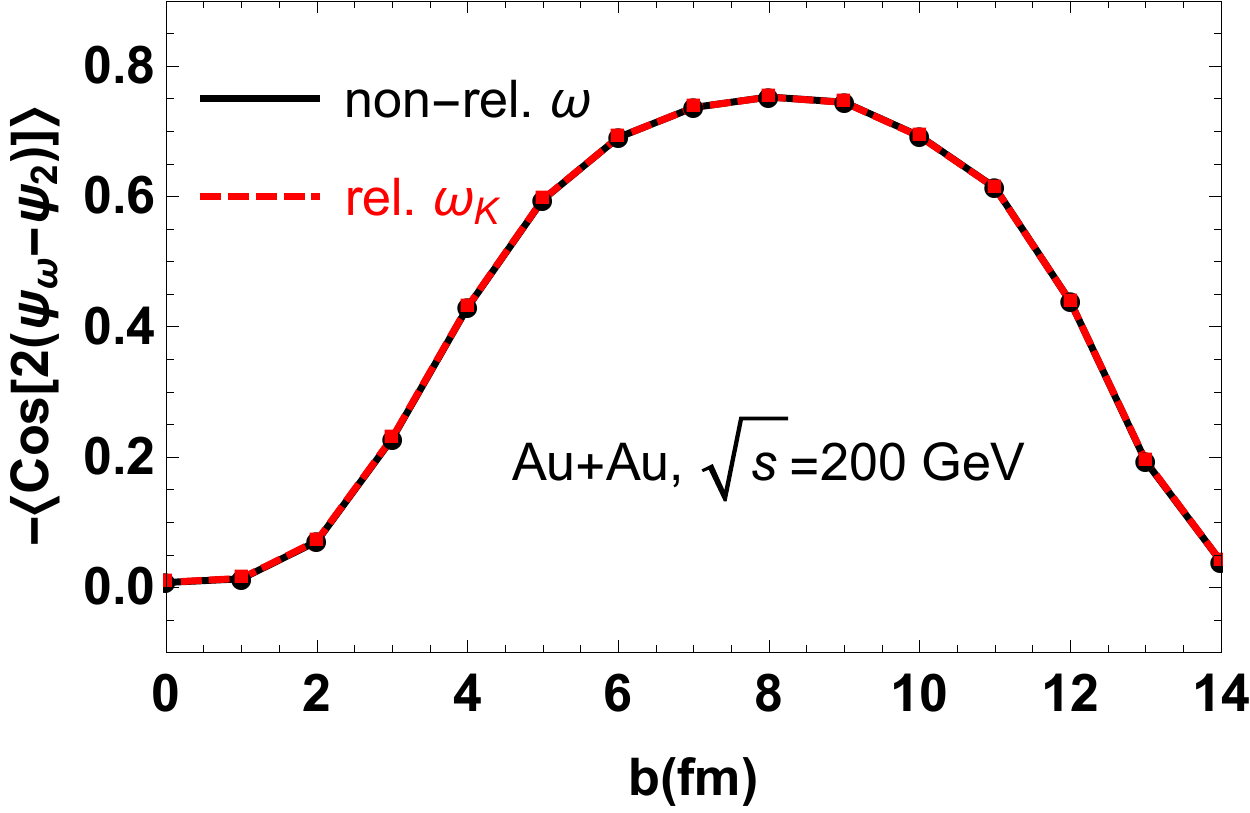}
\caption{The correlation between the direction of the vorticity $\psi_\omega$ and the second order participant plane $\psi_2$ \cite{Deng:2016gyh}.}
\label{cordepen}
\end{center}
\end{figure}
Geometrically, it is expected that the direction of the vorticity should be perpendicular to the reaction plane. However, this is true only at the optical limit or after event average. In reality, the nucleons in the nucleus are not static but always move from time to time, leading to the event-by-event fluctuation at the moment of collisions. Such event-by-event fluctuations can smear the direction of the vorticity from being perfectly perpendicular to the reaction plane. To quantify this effect, one can study the azimuthal angle correlation between the vorticity and the participant plane (which can describe the overlapping region more accurately than the reaction plane), $\langle \cos[2(\psi_\omega-\psi_2)]\rangle$, where $\psi_\omega$ and $\psi_2$ denote the azimuthal angle of the vorticity and the participant plane of the second order respectively. The result is shown in \fig{cordepen}. We see that the correlation is significantly suppressed in the most central (due to the strong fluctuation in $\psi_\omega$) and most peripheral (due to the strong fluctuation in $\psi_2$) collisions. We note that the similar feature can also be observed in magnetic fields~\cite{Bloczynski:2012en,Bloczynski:2013mca}.

\begin{figure}[!htb]
\begin{center}
\includegraphics[width=5.45cm]{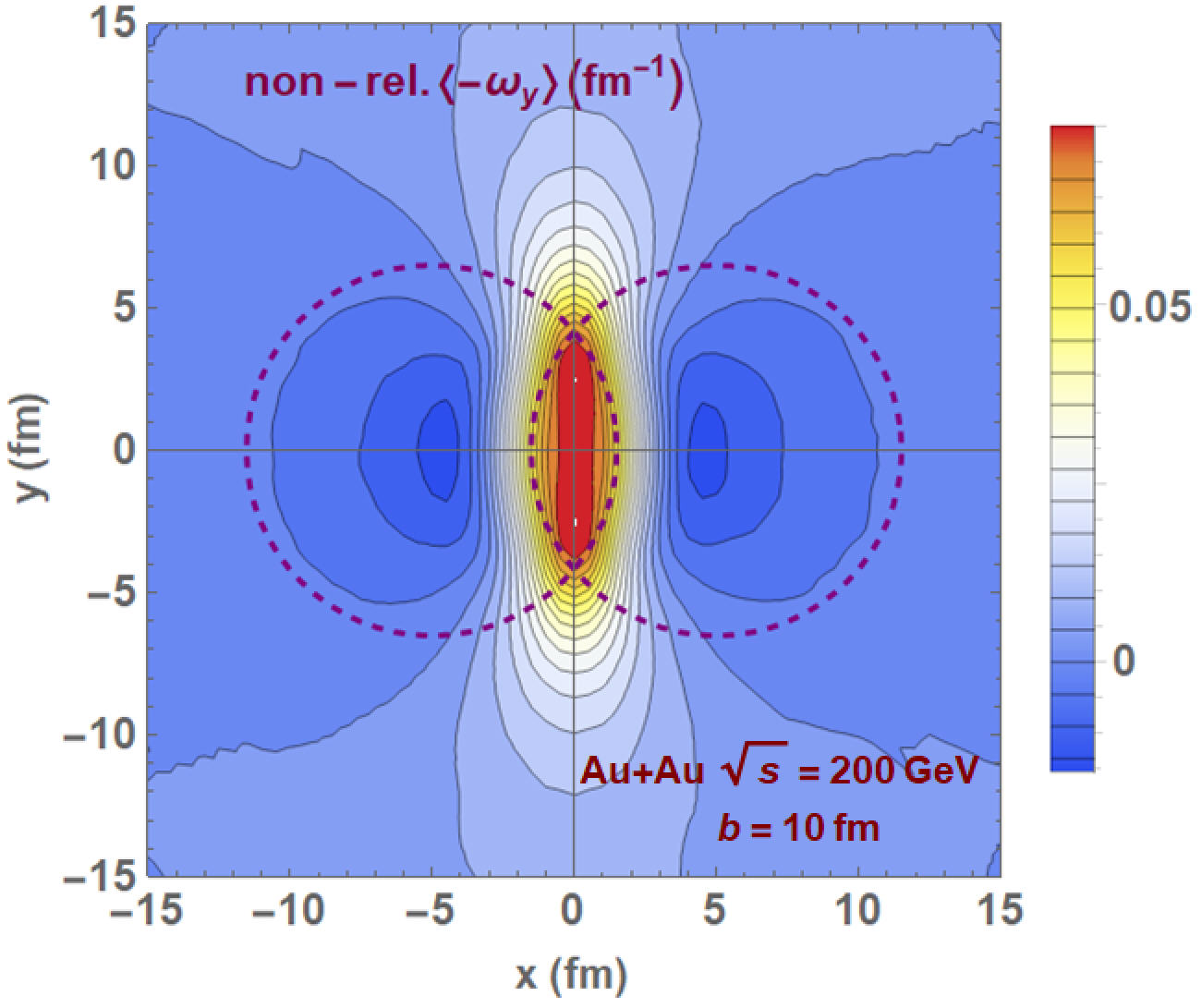}
\includegraphics[width=5.5cm]{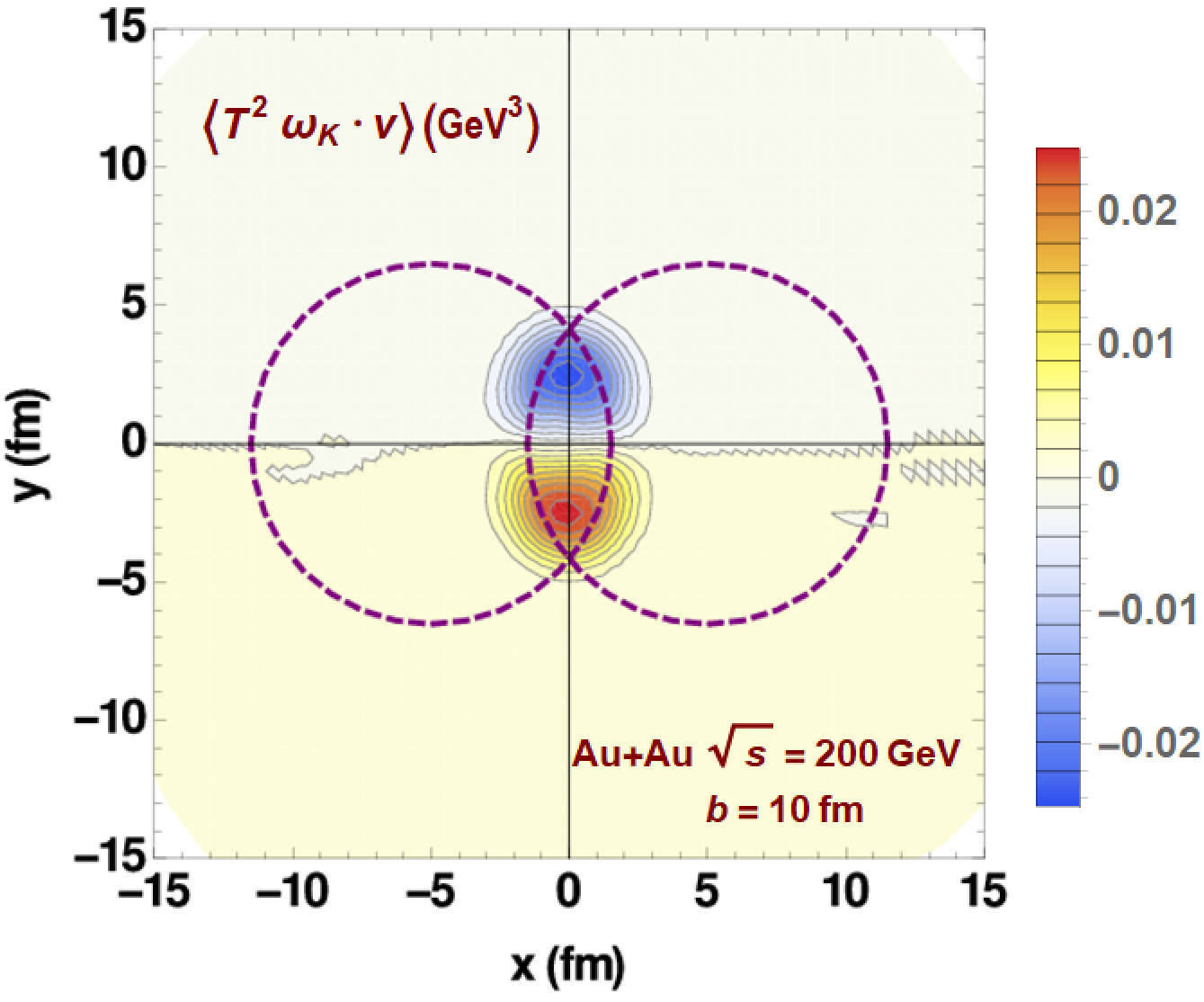}
\caption{The spatial distribution of the non-relativistic vorticity (left panel) and the helicity (right panel) in the transverse plane at $\eta=0$ for RHIC Au + Au collisions at $\sqrt{s}=200$ GeV~\cite{Deng:2016gyh}. See also discussion around \eq{eq:rheli}.}
\label{spadep}
\end{center}
\end{figure}
\textit{Spatial distribution.}
The vorticity is inhomogeneous in the transverse plane (the $x$-$y$ plane in \fig{frame}). As seen in \fig{spadep} (left panel), the non-relativistic vorticity varies more steeply along the $x$ direction in accordance with the elliptic shape of the overlapping region. The event avergage of the helicity field $h_T^0\equiv (1/2) T^2 \vec v\cdot\vec\nabla\times\vec v$ as defined below \eq{eq:rheli} is depicted in \fig{spadep} (right panel). Clearly, the reaction plane separates the region with the positive helicity from that with the negative helicity, due simply to the fact that $\langle v_y\rangle$ changes its sign across the reaction plane while $\langle\omega_y\rangle$ does not change sign. We note that a similar feature also exists for the electromagnetic helicity $\langle{\vec E\cdot\vec B}\rangle$ in heavy ion collisions~\cite{Deng:2012pc}.

\textit{Time evolution.}
In the hot quark-gluon medium, the fluid velocity evolves in time, so does the vorticity. Understanding the time evolution of the vorticity is also important for understanding vorticity-driven effects such as spin polarization. The results for the non-relativistic vorticity as functions of time in an AMPT simulation are presented in \fig{tdepen}. We see that at very early stage
$-\langle\bar{\omega}_y\rangle$ (in Ref. \cite{Jiang:2016woz}, the spacial average is weighted by the inertia moment) briefly increases with time which is probably due to a decrease of inertia moment by parton scatterings before the transverse
radial expansion is developed. After reaching a maximum value at $\sim 1$ fm, $-\langle\bar{\omega}_y\rangle$ follows a steady decrease with time because of the system's expansion.

To understand how the system's expansion brings the vorticity down, we can consider the dissipation equation for the non-relativistic vorticity,
\begin{eqnarray}
\label{eq:voreq}
\frac{\partial\vec \omega}{\partial t}=\vec\nabla\times(\vec v\times\vec\omega) +\nu\nabla^2\vec\omega,
\end{eqnarray}
where $\nu=\eta/(\varepsilon+P)=\eta/(sT)$ is the kinematic shear viscosity with $\eta$ being the shear viscosity and $s$ the entropy density. Thus, the change of the vorticity can be driven by either the fluid flow (the first term on the right-hand side) or by the viscous damping (the second term on the the right-hand side). The ratio of the two terms can be characterized by the Reynolds number $Re=UL/\nu$ with $U$ and $L$ being the characteristic velocity and system size respectively. If $Re\ll 1$, the second term dominates and the vorticity is damped by the shear viscosity with a time scale $t_\omega\sim L^2/(4\nu)$. If $Re\gg 1$, the first term in \eq{eq:voreq} dominates and the vortex flux is nearly frozen in the fluid (see the discussion in Sec.~\ref{subsec:nonrel} about the Helmholtz-Kelvin theorem). In this case, the vorticity decreases due to the system's expansion. Considering Au + Au collisions at
$\sqrt{s}=200$ GeV as an example. Typically, we can assume $U\sim 0.1 - 1$, $L \sim 5$ fm, $T\sim 300$ MeV, and $\eta/s\sim 1/(4\pi)$ for the strongly coupled QGP, then we have $Re\sim 10 - 100$. Thus, the vorticity decays as shown in \fig{tdepen} mainly due to the system's expansion, see Ref~\cite{Jiang:2016woz,Deng:2016gyh} for more discussions.

\subsection{Results for thermal vorticity}
\label{sec:num:thv}
The thermal vorticity (\ref{def:thv}) can be decomposed into the part
proportional to the kinematic vorticity and the part related to temperature gradients,
\begin{eqnarray}
\label{eq:thvsk}
\varpi_{\mu\nu}\equiv\omega^\beta_{\mu\nu}=\beta\omega^{\rm K}_{\mu\nu}+u_{[\nu}\partial_{\mu]}\beta,
\end{eqnarray}
where $[\cdots]$ means anti-symmetrization of indices. Note that in this sub-section we will use $\varpi$ to denote the thermal vorticity in order to be consistent with the traditional notation widely used in literature. Thus, in many aspects, the thermal vorticity behaves similarly to the kinematic vorticity. But the difference between two vorticities becomes significant when the temperature gradient is large.

\textit{Time evolution.}
\begin{figure}[!htb]
\begin{center}
\includegraphics[width=5cm]{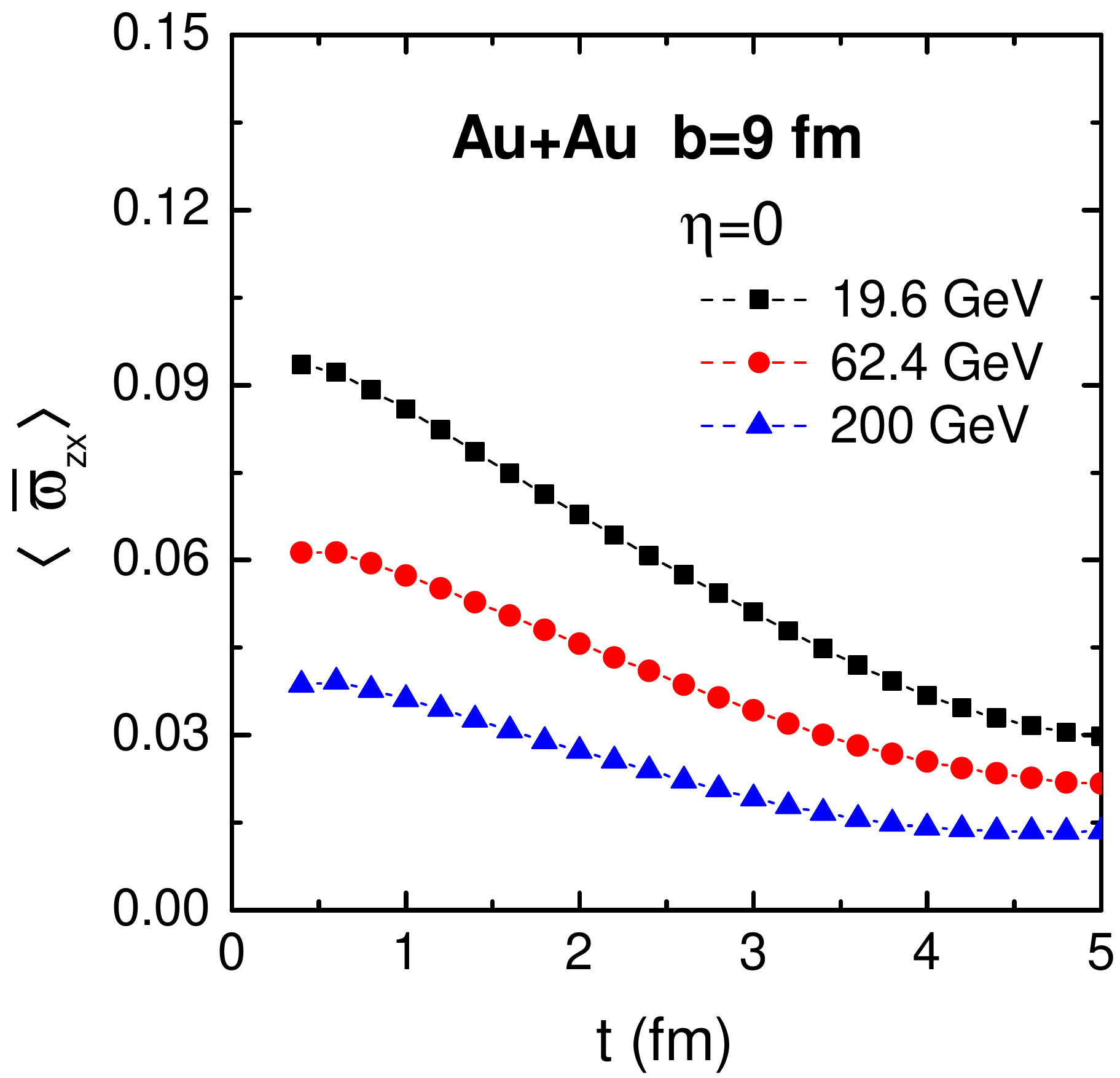}
\caption{The time evolution of the $zx$-component of the thermal vorticity at spacetime rapidity $\eta=0$ and impact parameter $b=9$ fm for different collision energies. The figure is taken from Ref.~\cite{Wei:2018zfb}.}
\label{th-t-depen}
\end{center}
\end{figure}
In \fig{th-t-depen}, we show the $zx$-component of the thermal vorticity in Au + Au collisions at $\eta=0$, $b=9$ fm and $\sqrt{s}=19.6, 62.4, 200$ GeV. Here, the thermal vorticity is averaged over the transverse plane first (weighted by the energy density and indicated by an overline) and then over collision events (indicated by $\langle\cdots\rangle$). Comparing with \fig{tdepen}, except for a very short early time, the time evolution of the thermal vorticity is similar to the kinematic vorticity, so is the energy dependence: both the thermal and kinematic vorticity decrease with $\sqrt{s}$. This can be understood from the fact that at higher collision energies both terms in \eq{eq:thvsk} become smaller at $\eta=0$ as two colliding nuclei become more transparent to each other and make the mid-rapidity region more boost-invariant.

\textit{Spatial distribution.}
\begin{figure}[!htb]
\begin{center}
\includegraphics[width=5cm]{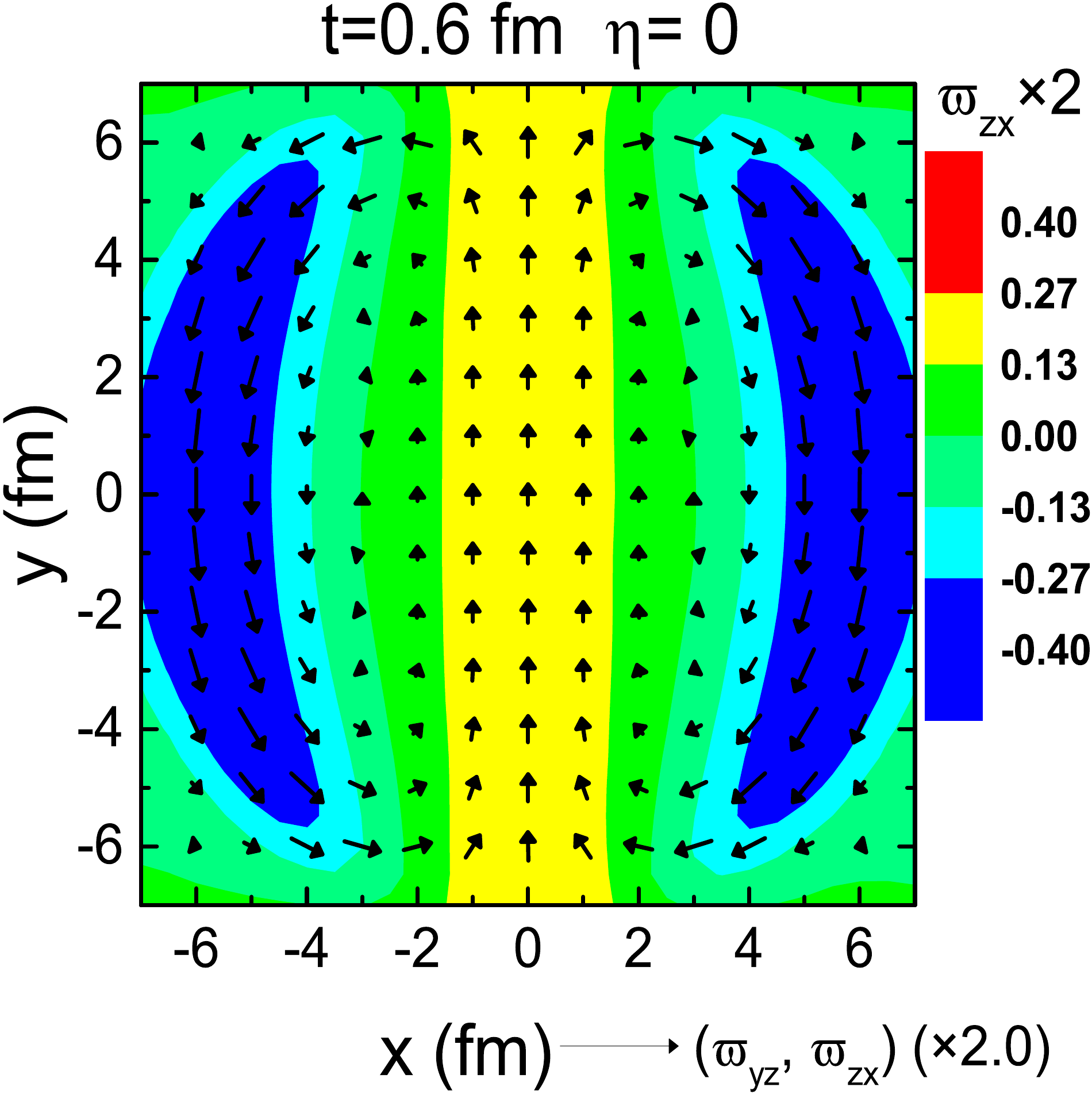}$\;\;\;\;\;\;\;\;\;\;\;$
\includegraphics[width=5cm]{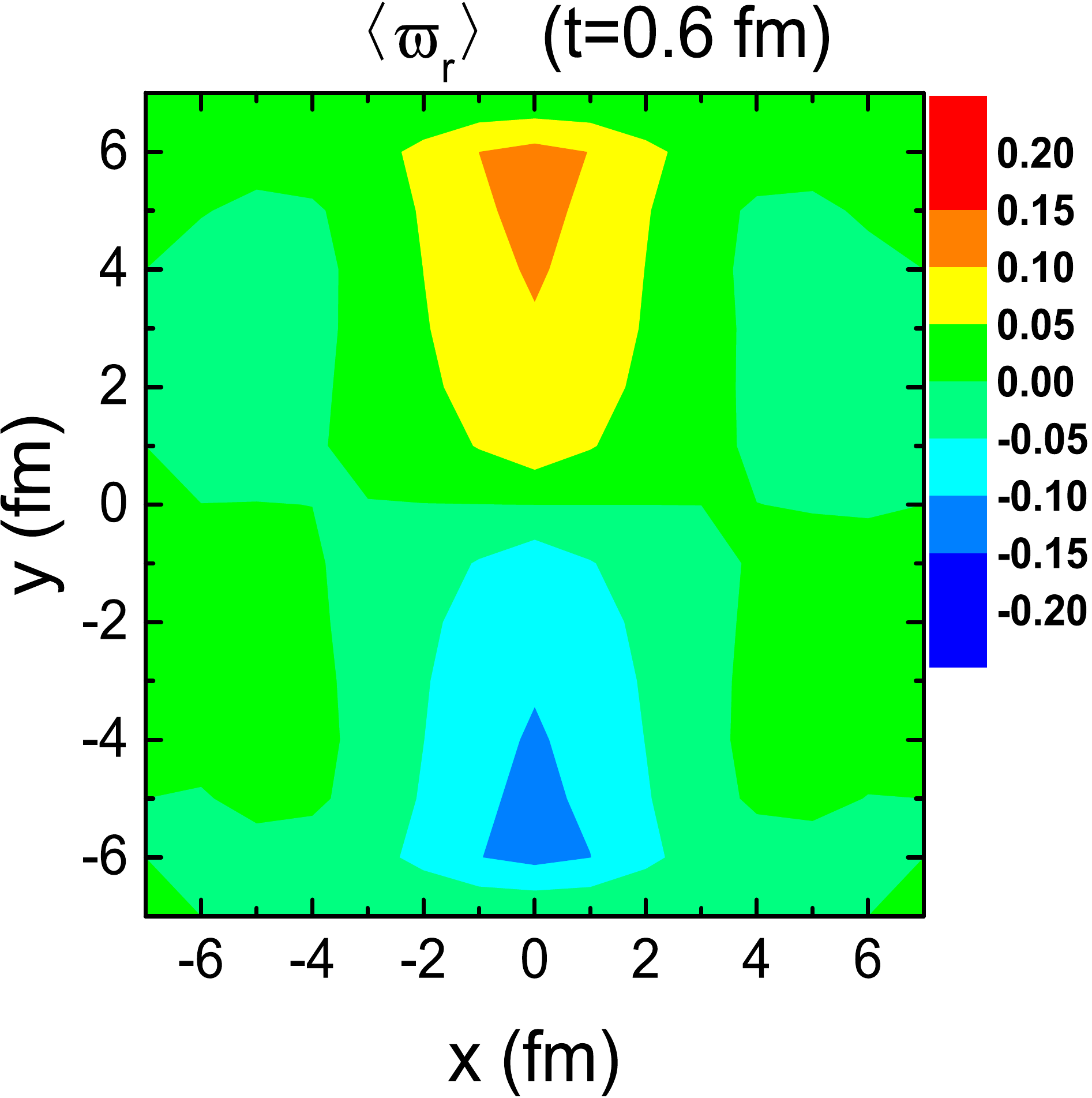}
\caption{The distribution of the event-averaged thermal vorticity on the transverse plane at $t=0.6$ fm,  $\eta=0$ and $\sqrt{s}=19.6$ GeV  for Au + Au collisions, averaged over the centrality range 20-50\%. Left panel: arrows represent $\langle\vec\varpi_\perp\rangle=(\langle\varpi_{yz}\rangle, \langle\varpi_{zx}\rangle)$ and colors represent the magnitude of $\langle\varpi_{zx}\rangle$. Right panel: the radial thermal vorticity $\langle\varpi_r\rangle=\hat{\vec r}\cdot\langle\vec\varpi_\perp\rangle$. The figures are taken from Ref.~\cite{Wei:2018zfb}.}
\label{th-spa-1}
\end{center}
\end{figure}
In the left panel of \fig{th-spa-1} we show the spatial distribution of the event-averaged thermal vorticity $\vec\varpi_\perp=(\varpi_{yz}, \varpi_{zx})$ on the transverse plane at $\eta=0$. We take $t=0.6$ fm for the Au + Au collisions at $\sqrt{s}=19.6$ GeV as an example. The arrows represent $\langle\vec\varpi_\perp\rangle$ and colors represent the magnitude of $\langle\varpi_{zx}\rangle$. We see two vorticity loops associated with the motion of the participant nucleons in the projectile and target nucleus, respectively. The right panel shows the radial component of $\langle\vec\varpi_\perp\rangle$ and a clear sign separation by the reaction plane is observed.

\begin{figure}[!htb]
\begin{center}
\includegraphics[width=4.6cm]{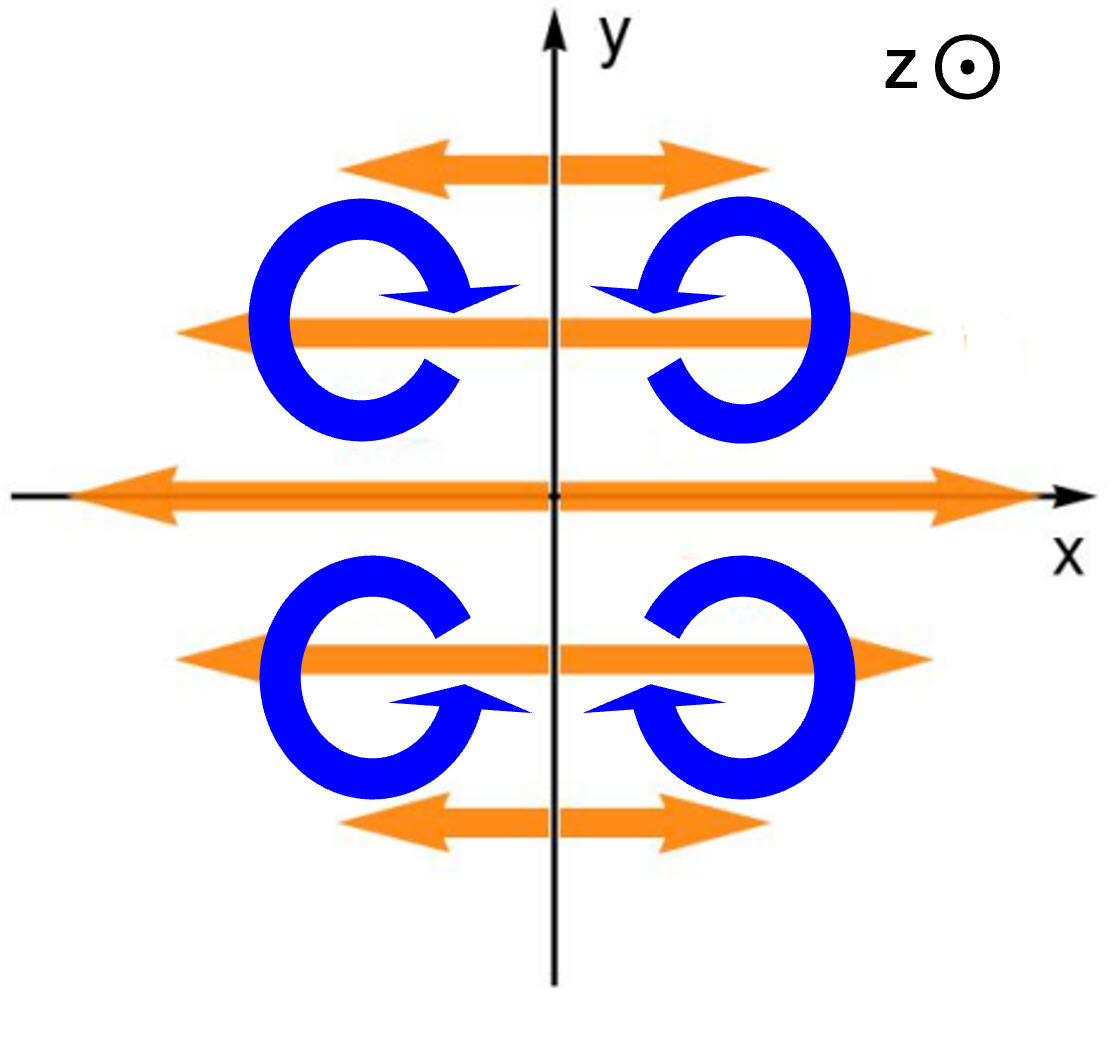}$\;\;\;\;\;\;\;\;\;\;\;$
\includegraphics[width=5cm]{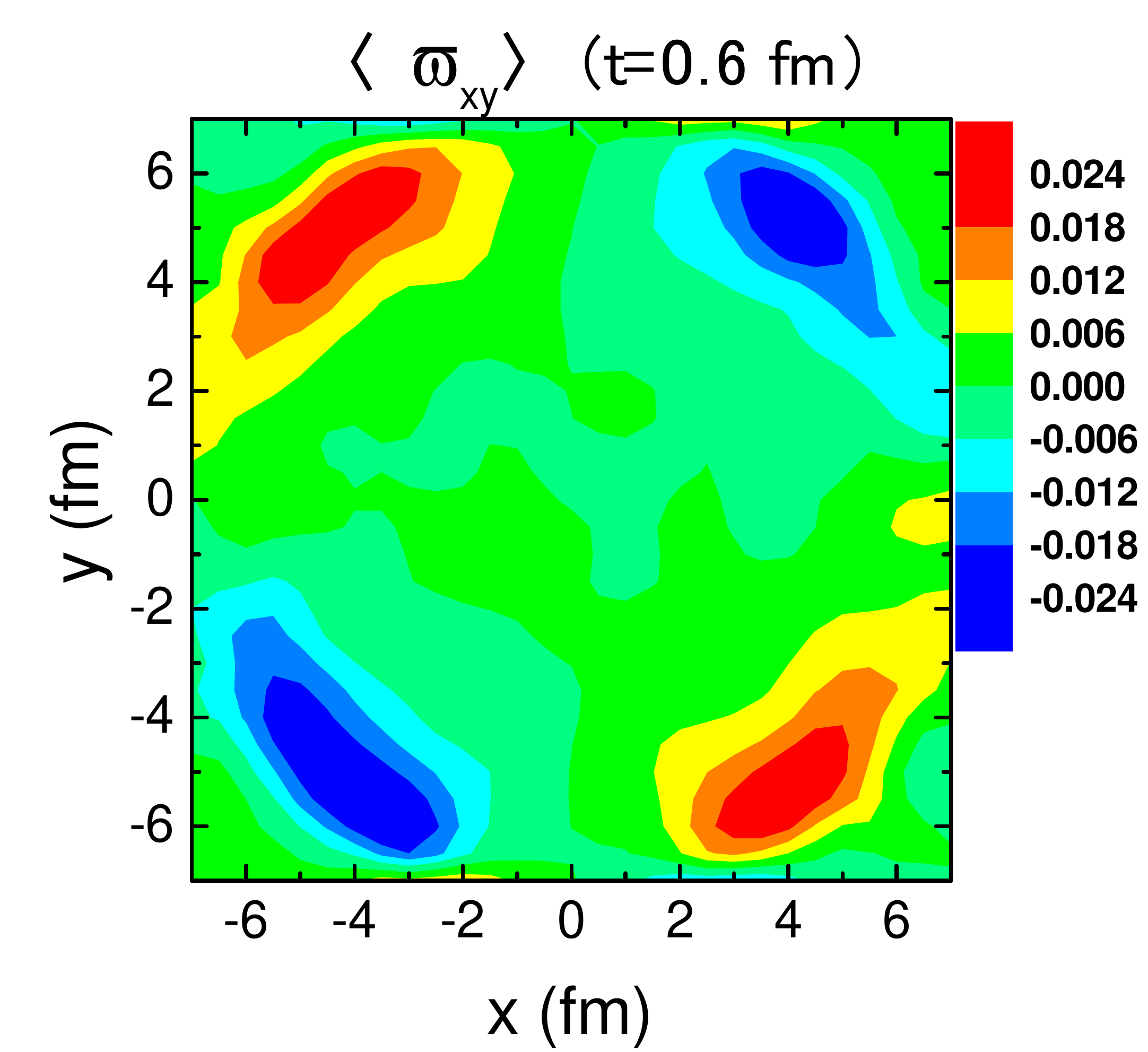}
\caption{Left panel: illustration of an anisotropic expansion of the fireball in the transverse plane in non-central collisions. Such a flow profile represents a positive elliptic flow $v_2$ and a quadrupolar distribution of the longitudinal kinematic vorticity \eq{defvor1} in the transverse plane. Right panel: the longitudinal component of the thermal vorticity distributed in the transverse plane at $t=0.6$ fm, $\eta=0$ and $\sqrt{s}=19.6$ GeV in Au + Au collisions. The results are obtained by averaging over events in $20-50$\% centrality. A remarkable difference between the left and right panel is the sign difference of the longitudinal vorticity in each quadrant. The figure is taken from Ref.~\cite{Wei:2018zfb}.}
\label{th-spa-long}
\end{center}
\end{figure}
As we have already discussed, the source of vorticty is multifold. The inhomogeneous expansion of the fireball serves as a good generator of the vorticity. To see this more clearly, let us consider a non-central collision and parameterize its velocity profile at a given moment as
\begin{eqnarray}
\label{expan}
v_r&\sim&\bar{v}_r(r,z)\left[1+2c_r\cos(2\phi)\right],\nonumber\\
v_z&\sim&\bar{v}_z(r,z)\left[1+2c_z\cos(2\phi)\right],\nonumber\\
v_\phi&\sim&2c_\phi \bar{v}_\phi(r,z) \sin(2\phi),
\end{eqnarray}
where $r$, $z$ and $\phi$ are the radial, longitudinal and azimuthal coordinates respectively, and $c_r$, $c_z$ and $c_\phi$ characterize the eccentricity in $v_r, v_z$ and $v_\phi$ respectively. For high-energy collisions, the expansion respects approximately a $z\rightarrow -z$ reflection symmetry which requires that $\bar{v}_r(r,z)=\bar{v}_r (r,-z)$, $\bar{v}_z(r,z)=-\bar{v}_z (r,-z)$, and $\bar{v}_\phi(r,z)=\bar{v}_\phi (r,-z)$. Thus we find very interesting features in the non-relativistic kinematic vorticity field, $\vec\omega=(1/2)\vec\nabla\times\vec v$, from the velocity profile (\ref{expan}).

First, at mid-rapidity $\eta=0$ or $z=0$ in a non-central collision, the longitudinal non-relativistic kinematic vorticity $\omega_z$ can be nonzero while the transverse component $\omega_r$ and $\omega_\phi$ vanish. In particular, we have $\omega_z\sim \sin(2\phi)$ at mid-rapidity, featuring a quadrupole distribution as illustrated in the left panel of \fig{th-spa-long} ~\footnote{We note that the left panel of \fig{th-spa-long} is just for illustrative purpose, the real velocity profile is much more complicated including components which can contribute a positive $v_2$ but an opposite vortical structure to the one shown in the figure.}. Such a quadrupole structure in the non-relativistic vorticity field is a result of the positive elliptic flow $v_2$. Quite similarly, the longitudinal component of the thermal vorticity also shows a quadrupole structure in the transverse plane in the right panel of \fig{th-spa-long}, in which the results of $\varpi_{xy}$ in the transverse plane of Au + Au collisions at at $t=0.6$ fm, $\eta=0$ and $\sqrt{s}=19.6$ GeV are presented. Surprisingly, in each quadrant, the thermal vorticity $\varpi_{xy}$ has an opposite sign comparing with the non-relativistic vorticity $\omega_z$. This means that the contributions from acceleration and temperature gradient to the thermal vorticity are large and outperform that from the velocity gradient.

\begin{figure}[!htb]
\begin{center}
\includegraphics[width=5.8cm]{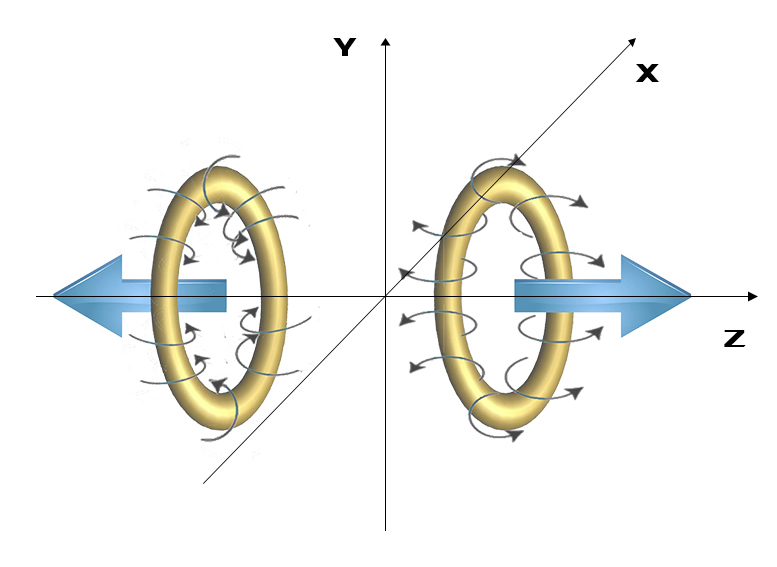}$\;\;\;\;$
\includegraphics[width=5cm]{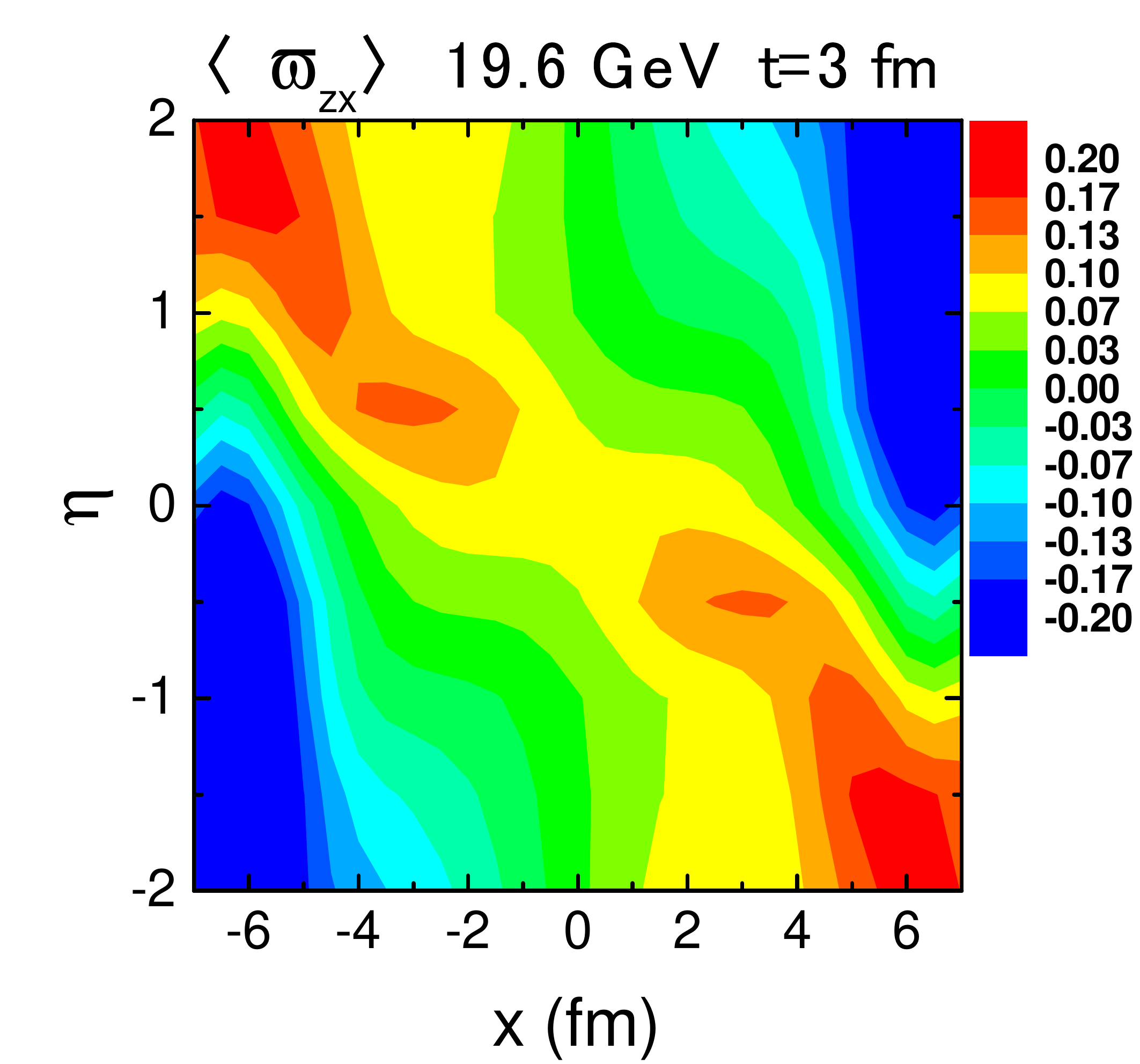}
\includegraphics[width=4.8cm]{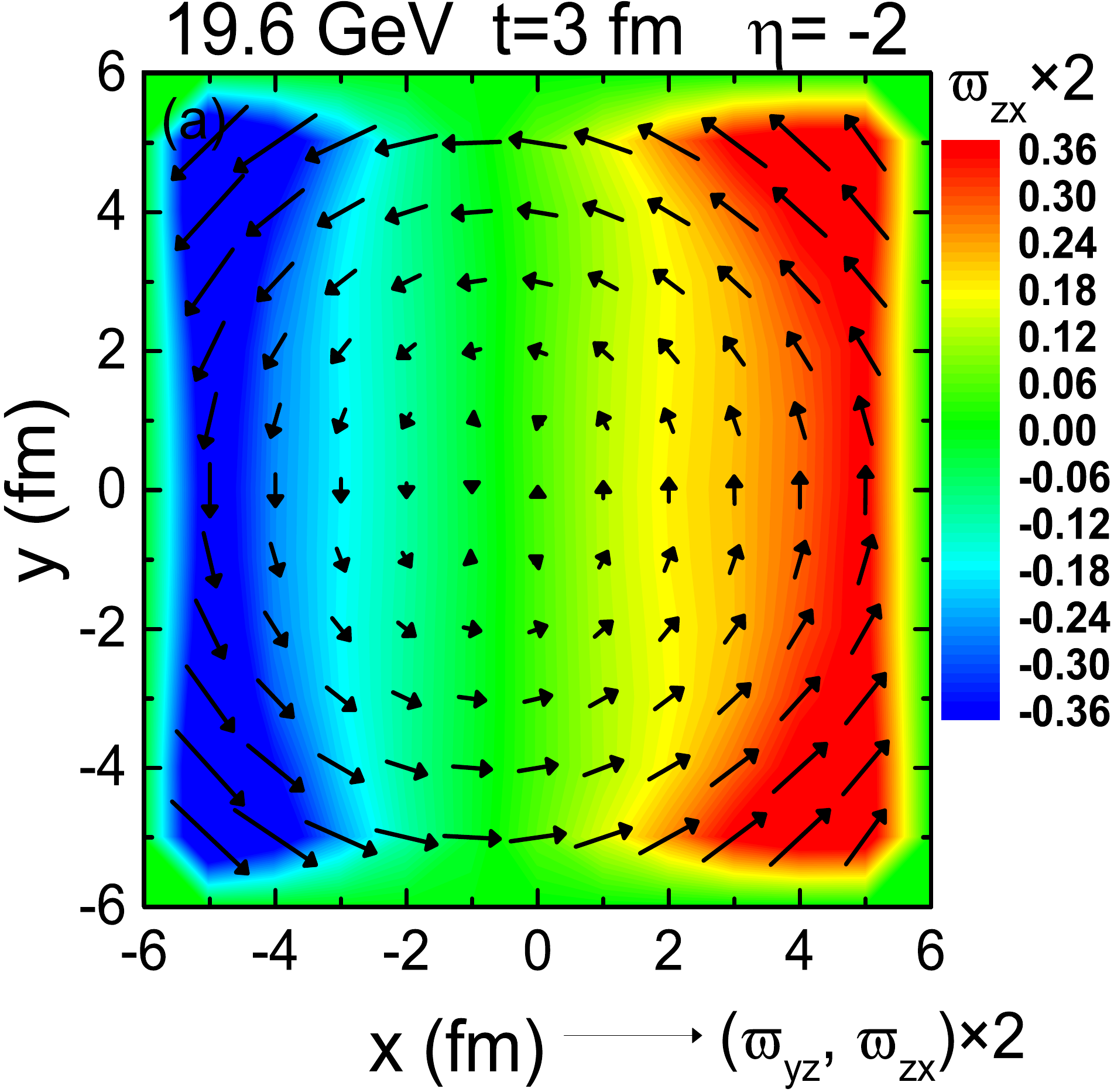}$\;\;\;\;\;\;\;\;\;\;\;\;\;\;\;\;\;\;\;$
\includegraphics[width=4.8cm]{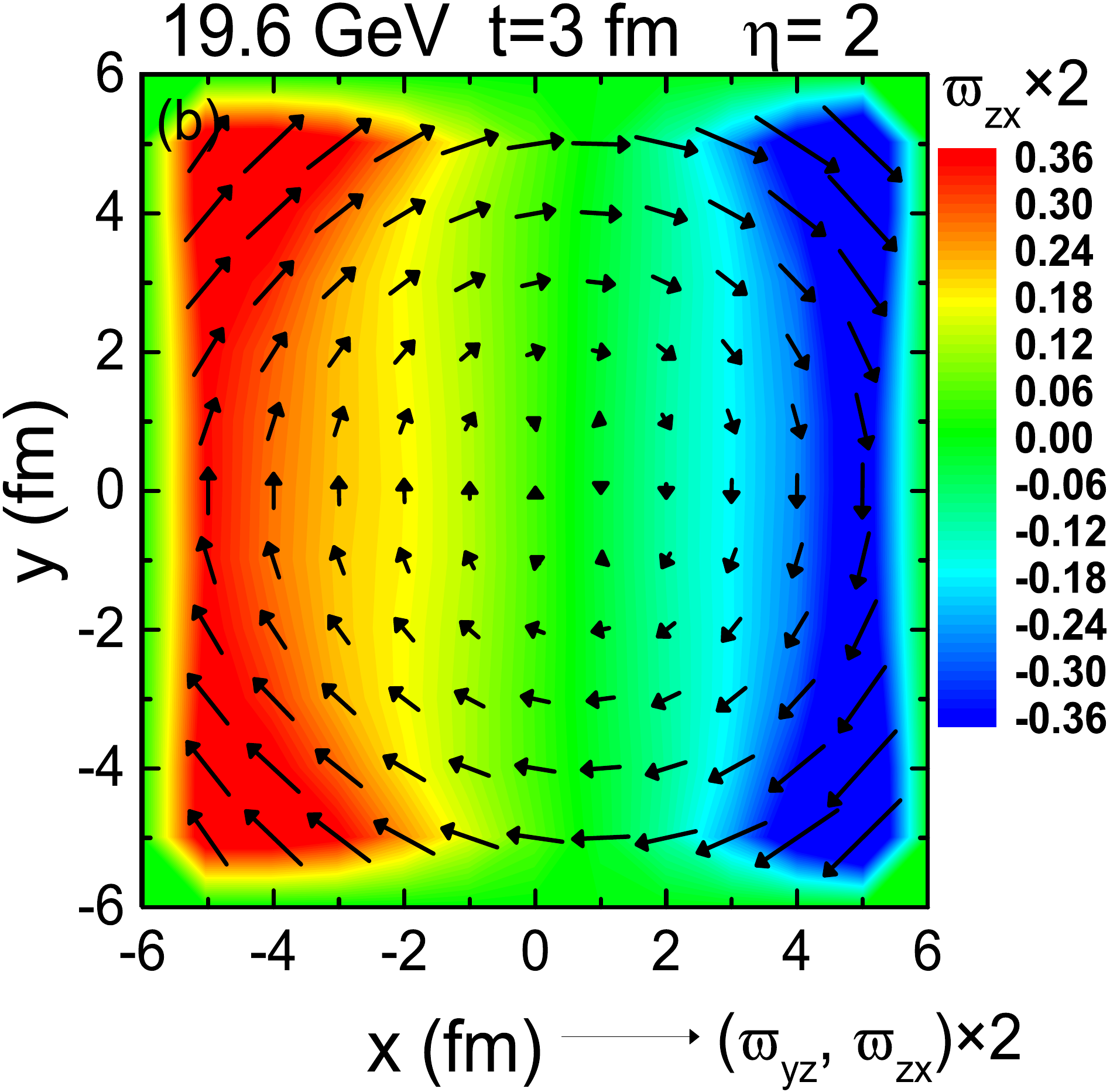}
\caption{Upper-left: the illustration of the smoke-loop type vortices due to the fast longitudinal expansion. Note that the radial expansion inhomogeneous in $z$ or $\eta$ direction results in similar vortices. Upper-right: the distribution of event-averaged thermal vorticity in the reaction plane (the $x-\eta$ plane) for Au + Au collisions at $\sqrt{s}=19.6$ GeV. Lower-left and lower-right: the vector plot for the thermal vorticity projected to the transverse plane at spacetime rapidity $\eta = -2,2$ for Au + Au collisions at $\sqrt{s}=19.6$ GeV averaged over events in 20-50\% centrality range. The background color represents the magnitude and sign of $\varpi_{zx}$. The figures are from Ref.~\cite{Wei:2018zfb}.}
\label{th-spa-eta}
\end{center}
\end{figure}

Second, at finite rapidity, all three components of $\vec\omega$ can be finite and the transverse vorticity is dominated by the $\phi$ component. The origin of this $\phi$-directed vortex is similar to the onset of the smoke-loop vortex as illustrated in \fig{th-spa-eta} (upper-left). More precisely, $\omega_\phi\sim(1/2)[\partial\bar{v}_r/\partial z-\partial\bar{v}_z/\partial r]$ changes sign under the relection transformation $z\rightarrow -z$ or $\eta\rightarrow -\eta$, such a behavior exists in non-central as well as central collisions. In positive rapidity region $\eta>0$, the first term in $\omega_\phi$ is usually negative while the second term is positive, so the direction of the $\phi$-directed vortex depends on the relative strength of two terms. Similar smoke-loop pattern for the thermal vorticity also exists, see the lower panels of \fig{th-spa-eta}. The projection to the reaction plane forms a quadrupole structure for $\varpi_{zx}$ as shown in the upper-right panel of \fig{th-spa-eta}. We will discuss how this intricate local vortical structure can be reflected in the spin polarization of $\Lambda$ hyperons in next section.


\section{$\Lambda$ polarization in heavy ion collisions}
\label{sec:pola:num}
An important consequence of the vorticity field is that particles with spin can be polarized. The detailed mechanism for such a spin polarization has been discussed in Sec.~\ref{sec:pola}. In this section, we review the numerical simulation based on transport models for the spin polarization of one specific hyperon, $\Lambda$ and its antiparticle, $\bar\Lambda$. The reason why the $\Lambda$ hyperon is chosen is that its weak decay $\Lambda\rightarrow p +\pi^-$ which violates the parity symmetr, so the daughter proton  emits preferentially along the spin direction of $\Lambda$ in its rest frame. More precisely, if $\vec P^*_\Lambda$ is the spin polarization of $\Lambda$ in its rest frame (hereafter, we will use an asterisk to indicate $\Lambda$'s rest frame), the angular distribution of the daughter protons is given by
\begin{eqnarray}
\label{lambdadecay}
\frac{1}{N_p}\frac{d N_p}{d\Omega^*}=\frac{1}{4\pi}\left(1+\alpha\hat{\vec p}^*\cdot\vec P^*_\Lambda\right),
\end{eqnarray}
where $\vec p^*$ is the momentum of the proton in the rest frame of $\Lambda$ (a hat over a vector denotes its unit vector), $\Omega^*$ is the solid angle of $\vec p^*$, and $\alpha\approx 0.642\pm 0.013$ is the decay constant. Thus, experimentally, one can extract $P^*_\Lambda$ by measuring $dN_p/d\Omega^*$ \cite{STAR:2017ckg,Abelev:2007zk,Siddique:2017ddr}. The above discussion applies equally well to $\bar\Lambda$ but with a negative decay constant $-\alpha$. Our purpose is to discuss the current theoretical understanding of $\vec P^*_\Lambda$ induced by the vorticity in heavy ion collisions. We note that the vorticity induced spin polarization can also lead to other interesting consequences like the spin alignment of vector mesons,~\cite{Liang:2004xn,Sheng:2019kmk,Sheng:2020ghv,Xia:2020tyd}, enhancement of the yield of hadrons with higher spin~\cite{Taya:2020sej}, spin-spin correlation~\cite{Pang:2016igs}, polarization of emitted photons~\cite{Ipp:2007ng}, which, however, will not be discussed.

The basic assumption that enables us to link the vorticity and $\Lambda$ spin polarization is the local equilibrium of the spin degree of freedom leading to the formula for a spin-$s$ fermion with mass $m$ and momentum $p^\mu$ produced at point $x$ \cite{Fang:2016vpj,Liu:2020flb,Becattini:2013fla,Becattini:2016gvu},
\begin{eqnarray}
\label{spin}
S^\mu(x,p)=-\frac{s(s+1)}{6m}(1-n_F)\epsilon^{\mu\nu\rho\sigma} p_\nu \varpi_{\rho\sigma}(x)+O(\varpi)^2,
\end{eqnarray}
where $n_F(p_0)$ is the Fermi-Dirac distribution function with $p_0=\sqrt{\vec p^2+m^2}$ being the energy of the fermion. We should note that this formula can be shown to be hold at global equilibrium, as we derived in Sec.~\ref{sec:pola}, but here we assume that it holds also at local equilibrium. For $\Lambda$ and $\bar{\Lambda}$, we have $s=1/2$. If the fermion mass is much larger than the temperature as in the case of $\Lambda$ and $\bar{\Lambda}$ produced in heavy ion collisions at RHIC and LHC energies, we can approximate $1-n_F\approx 1$. Using $S^{*\mu}=(0,{\vec S}^{*})$ to denote the spin vector in $\Lambda$'s rest frame, the Lorentz transformation from the laboratory frame gives
\begin{eqnarray}
\label{localspin}
{\vec S}^{*}={\vec S}-\frac{{\vec p}\cdot{\vec S}}{p_0(p_0+m)}{\vec p}.
\end{eqnarray}
Finally, the spin polarization of $\Lambda$ in the direction ${\vec n}$ is given by
\begin{eqnarray}
\label{def:pol}
P^*_n= \frac{1}{s}{\vec S}^{*}\cdot{\vec n}.
\end{eqnarray}
In the following, for simplicity, we will use $P_n$ to denote $P^*_n$ if there is no confusion. In a transport model like AMPT, \eqs{spin}{def:pol} are used to calculate the $\Lambda$ polarization.

\begin{figure}[!htb]
\begin{center}
\includegraphics[width=5.6cm]{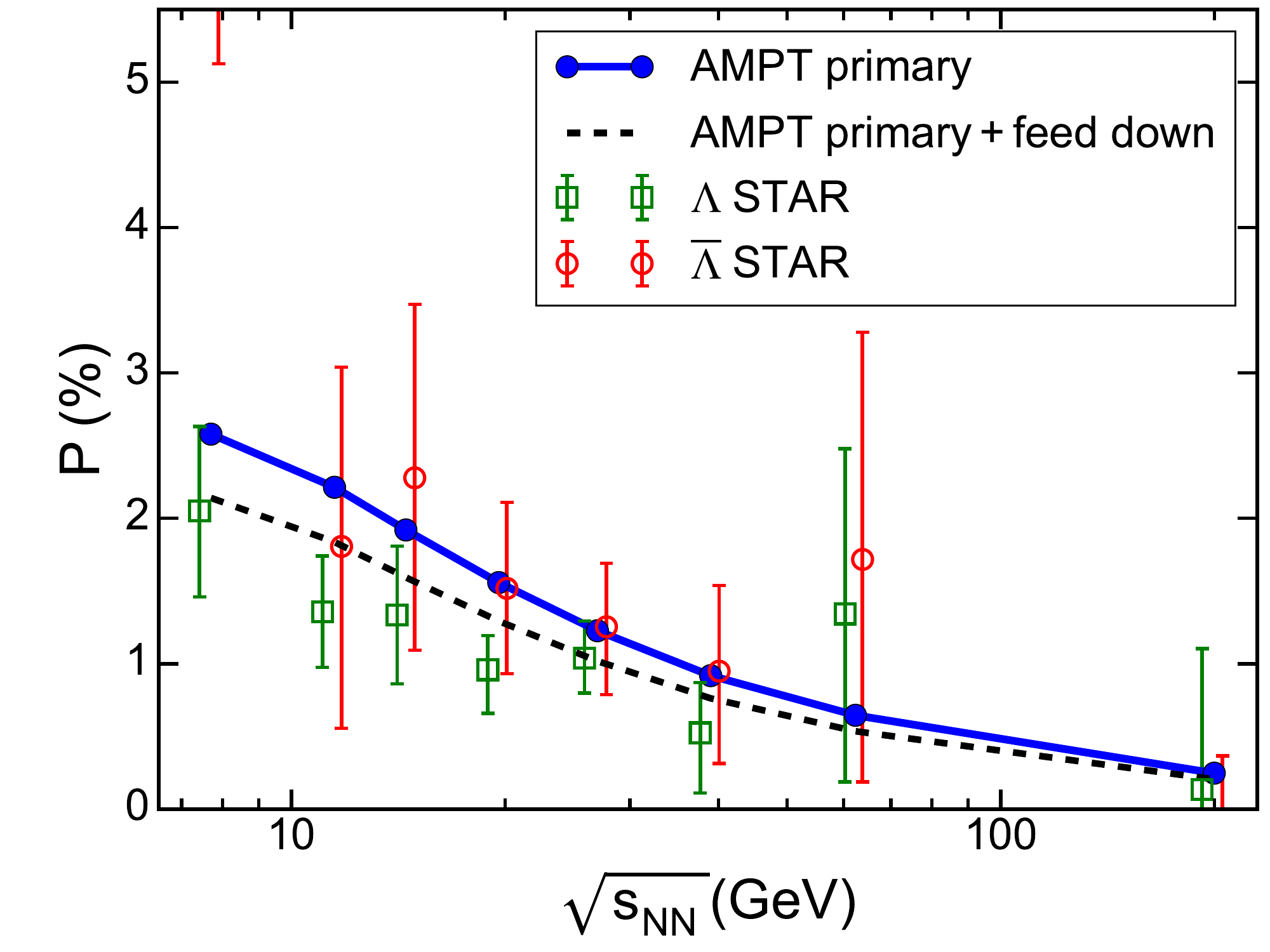}
\includegraphics[width=5.5cm]{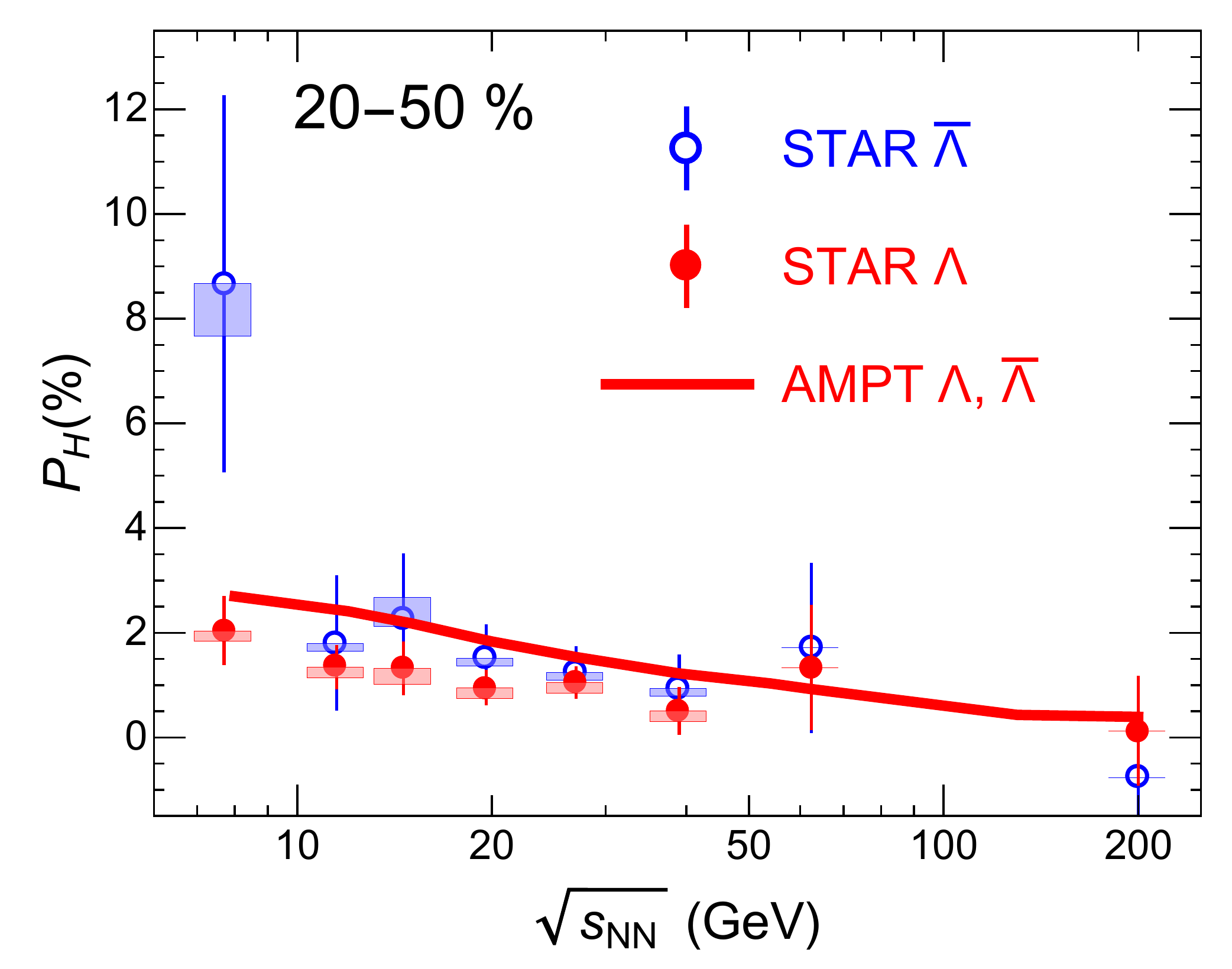}
\includegraphics[width=5cm]{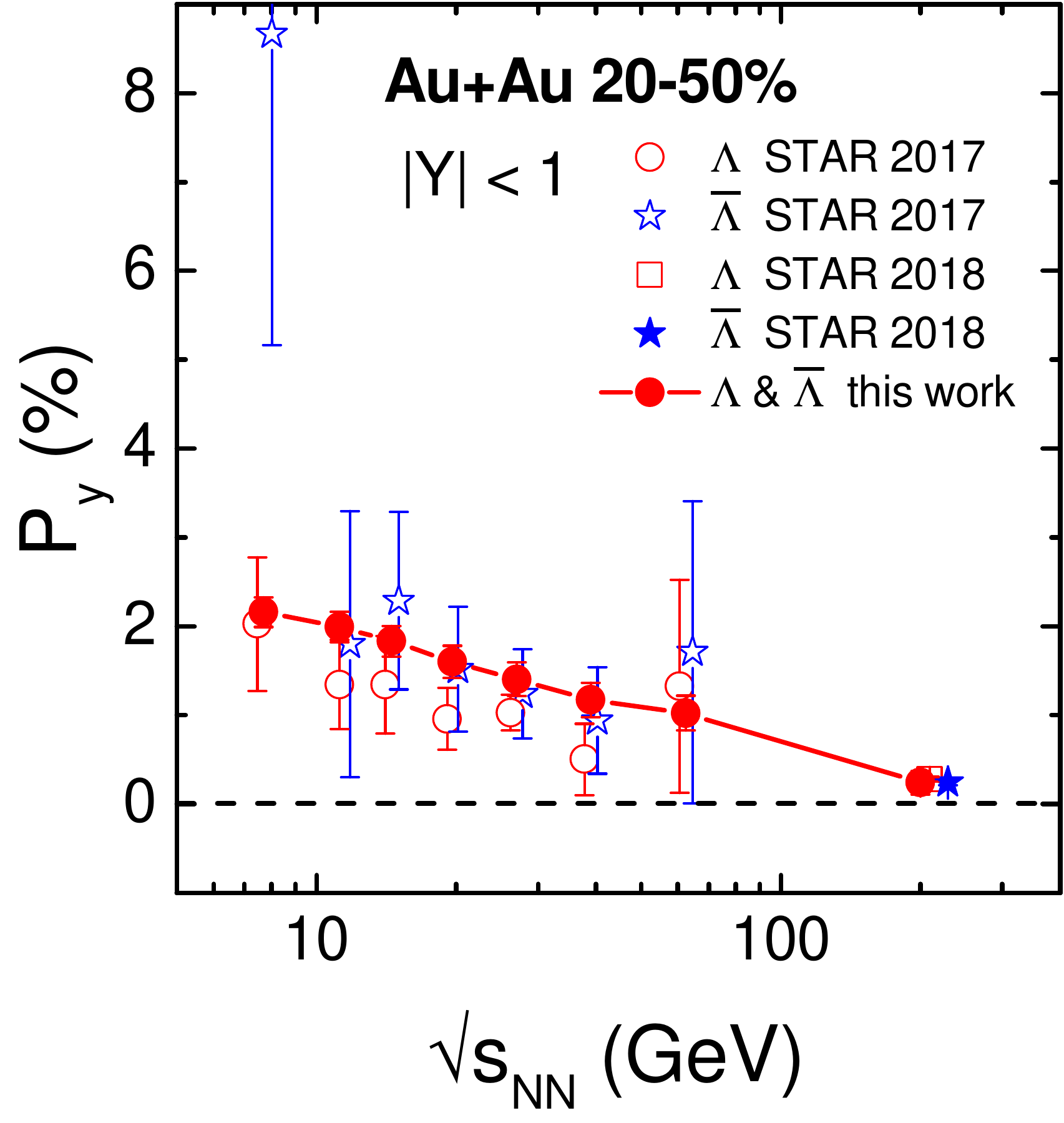}
\caption{The global $\Lambda$ and $\bar\Lambda$ spin polarization simulated in AMPT model with comparison to the experimental data. Shown are the polarization along y direction in 20-50\% centrality range of Au+Au collisions from different working groups which are consistent to each other~\cite{Li:2017slc,Shi:2017wpk,Wei:2018zfb}.}
\label{pol:global}
\end{center}
\end{figure}

\textit{The global polarization.} In the last few years, transport models such as AMPT have been
widely used in the study of the $\Lambda$ polarization. The results of various groups are consistent
to each other to a large extent. Here, we mainly show the results of Ref.~\cite{Li:2017slc,Wei:2018zfb,Shi:2017wpk}.
In \fig{pol:global}, theoretical results for the spin polarization of $\Lambda$ and $\bar\Lambda$ are compared
with experimental data. The simulations are done for Au + Au collisions in the centrality range $20-50\%$
and rapidity range $|Y|<1$.  We see very good agreement between numerical results and data, which
gives a strong support for the vorticity interpretation of the measured $\Lambda$ polarization.
We have three comments. (1) The simulations include only the polarization caused by the vorticity,
so there is no difference between $\Lambda$ and $\bar\Lambda$ in the calculation.
The data show a difference between $\Lambda$ and $\bar\Lambda$ although
the errors are are large. This is not fully understood. A possible source for such a
difference might be the magnetic field because $\Lambda$ and $\bar\Lambda$ have an opposite magnetic moment.
(2) The simulation given in \fig{pol:global} counts only the $\Lambda$ and $\bar\Lambda$
coming from hadronization of quarks in the AMPT model (called the primary or
primordial $\Lambda$ and $\bar\Lambda$). However, a big fraction ($\sim 80\%$) of
the measured $\Lambda$ and $\bar\Lambda$ hypersons are from the decay
of higher-lying hyperons such as $\Sigma^0$, $\Sigma^*$, $\Xi$, and $\Xi^*$.
However, such a feed-down contribution to the $\Lambda$ polarization is small:
it can reduce about $10-20\%$ of the spin polarization of primordial $\Lambda$'s \cite{Xia:2019fjf,Becattini:2019ntv}.
(3) Recently, HADES collaboration reported the measurement of the $\Lambda$ polarization
at $\sqrt{s}=2.4$ GeV which shows a nearly vanishing $P_y$~\cite{HADES:2019}. This means that the energy
dependence of $P_y$ at low energies is not monotonous. The AMPT model is not applicable
in such a low-energy region, and it is necessary to use other transport models,
such as UrQMD and IQMD,
to calculate the $\Lambda$ polarization at very low energy~\cite{Deng:2020ygd}.

\textit{Local polarization and polarization harmonics.}
The above analysis is for the integrated spin polarization over the azimuthal angle and rapidity and $p_T$ region, so is called the global polarization. As we have shown in Sec.~\ref{sec:num:thv}, the thermal vorticity has a nontrivial distribution in coordinate space, especially the quadrupole  structure shown in \fig{th-spa-long} and \fig{th-spa-eta}, leading to a nontrivial spin-polarization distribution in momentum space following Eq. (\ref{average-spin-vector}).

\begin{figure}[!htb]
\begin{center}
\includegraphics[height=5cm]{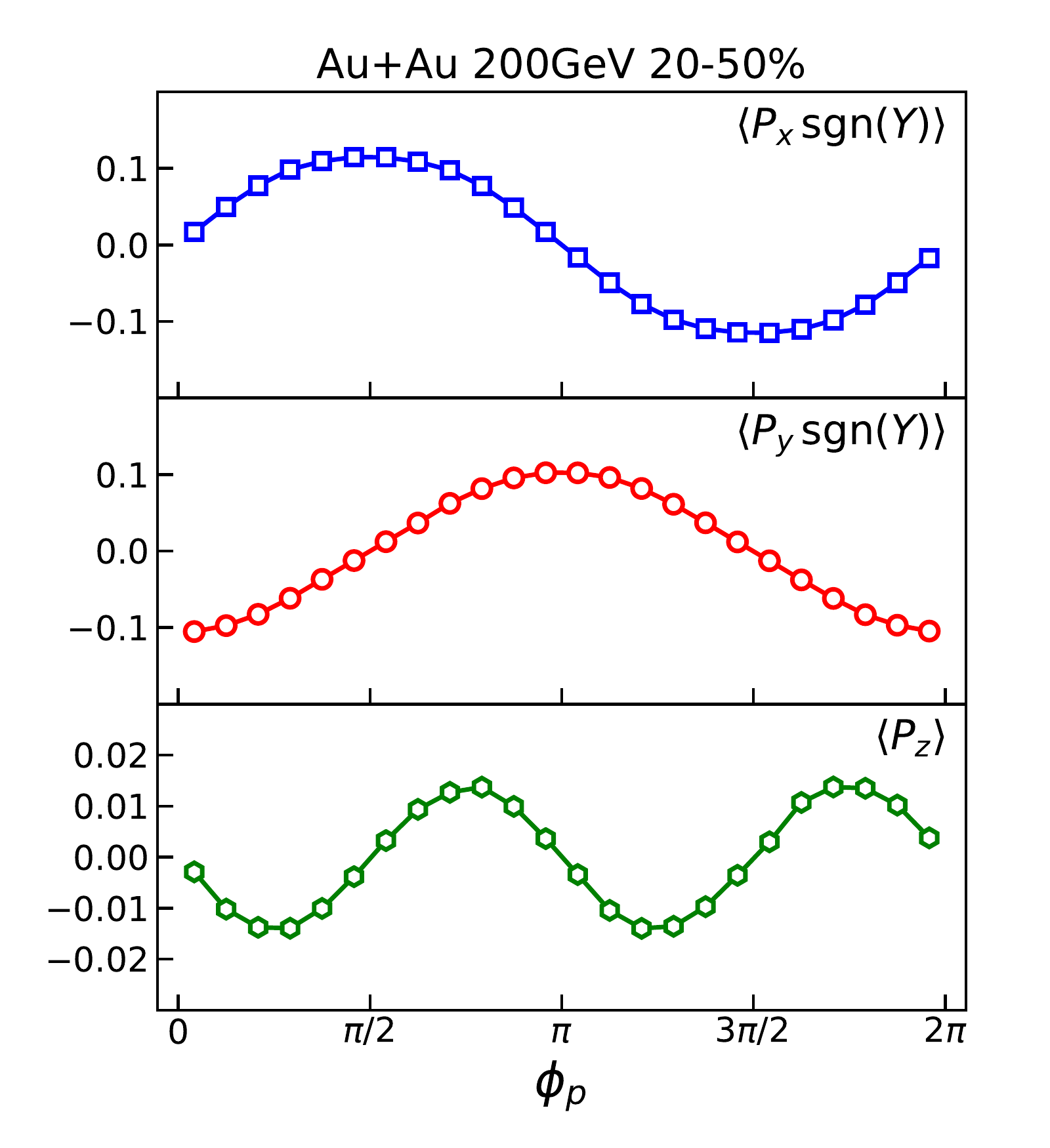}
\includegraphics[height=5cm]{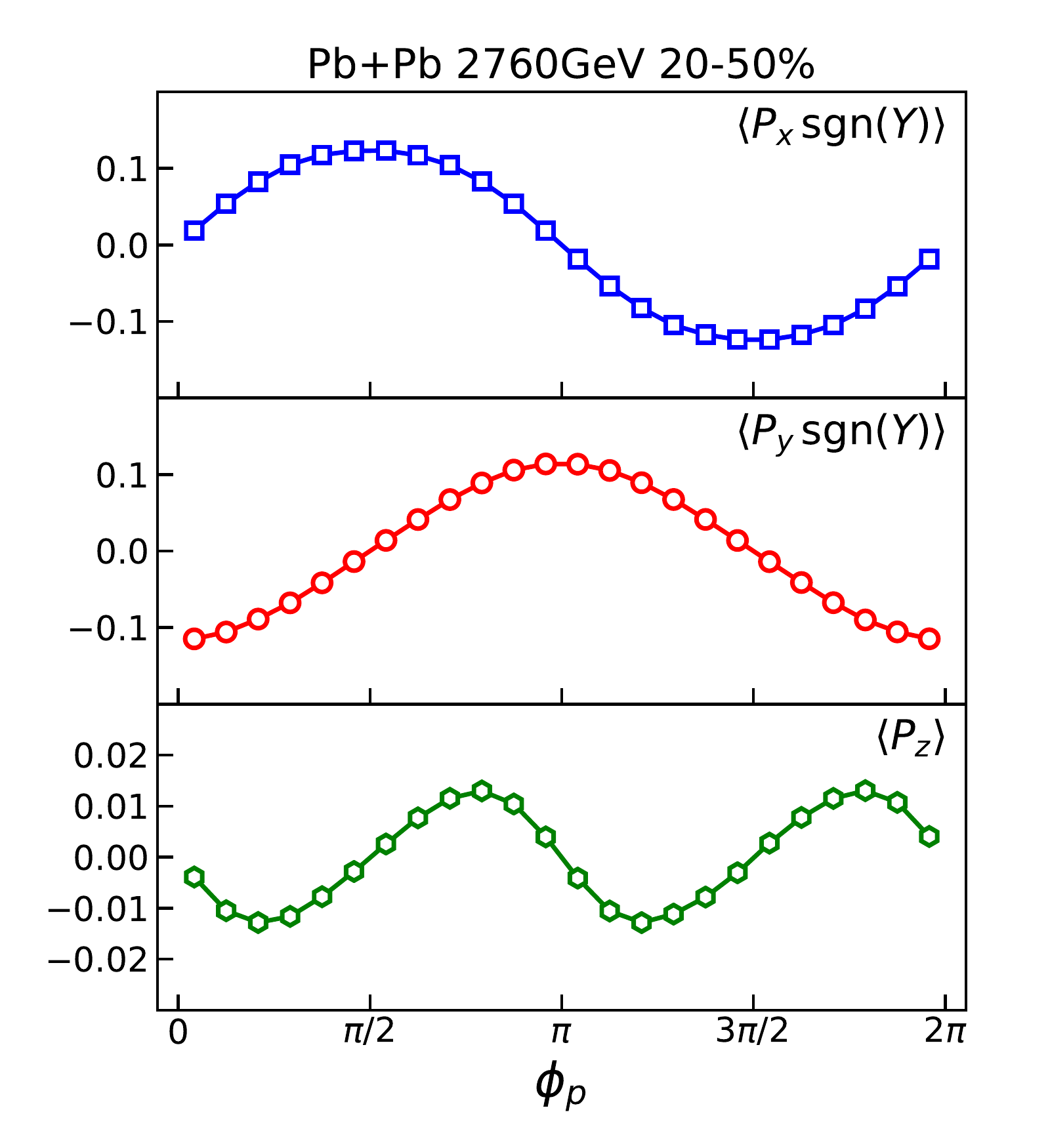}
\caption{The $\Lambda$ spin polarization as functions of $\Lambda$s' momentum azimuthal angle $\phi_p$ in 20--50\% central Au + Au
collisions at 200 GeV (left) and Pb + Pb collisions at 2760 GeV (right)~\cite{Xia:2018tes}.}
\label{pol:local-Xia}
\end{center}
\end{figure}

Here, we show the results from Ref.~\cite{Xia:2018tes}. In \fig{pol:local-Xia}, we present the $\Lambda$ spin polarization as functions of $\Lambda$s' momentum azimuthal angle $\phi_p$ for Au + Au collisions at 200 GeV (left) and Pb + Pb collisions at 2760 GeV (right). As illustrated in \fig{th-spa-eta}, the inhomogeneous expansion of the fireball can generate transverse vorticity loops, the directions of which are clockwise and counterclockwise in positive and negative rapidity regions, respectively. As a consequence, the transverse $\Lambda$ spin polarization $(P_x, P_y)$ should have a similar structure. To extract this effect, $P_x$ and $P_y$ are weighted by the sign of rapidity and then averaged in local azimuthal-angle bins. The results shown in the upper two panels in \fig{pol:local-Xia} present good harmonic behaviors $P_x\mathrm{sgn}(Y)\sim\sin(\phi_p)$ and $P_y\mathrm{sgn}(Y)\sim-\cos(\phi_p)$, which agree with the direction of the transverse vorticity loop. On the other hand, the longitudinal vorticity has a quadrupole structure on the transverse plane shown in \fig{th-spa-long}. Correspondingly, the longitudinal spin polarization $P_z$ shows a $-\sin(2\phi_p)$ behavior in the lowest panels in \fig{pol:local-Xia}.

\begin{figure}[!htb]
\begin{center}
\includegraphics[height=5cm]{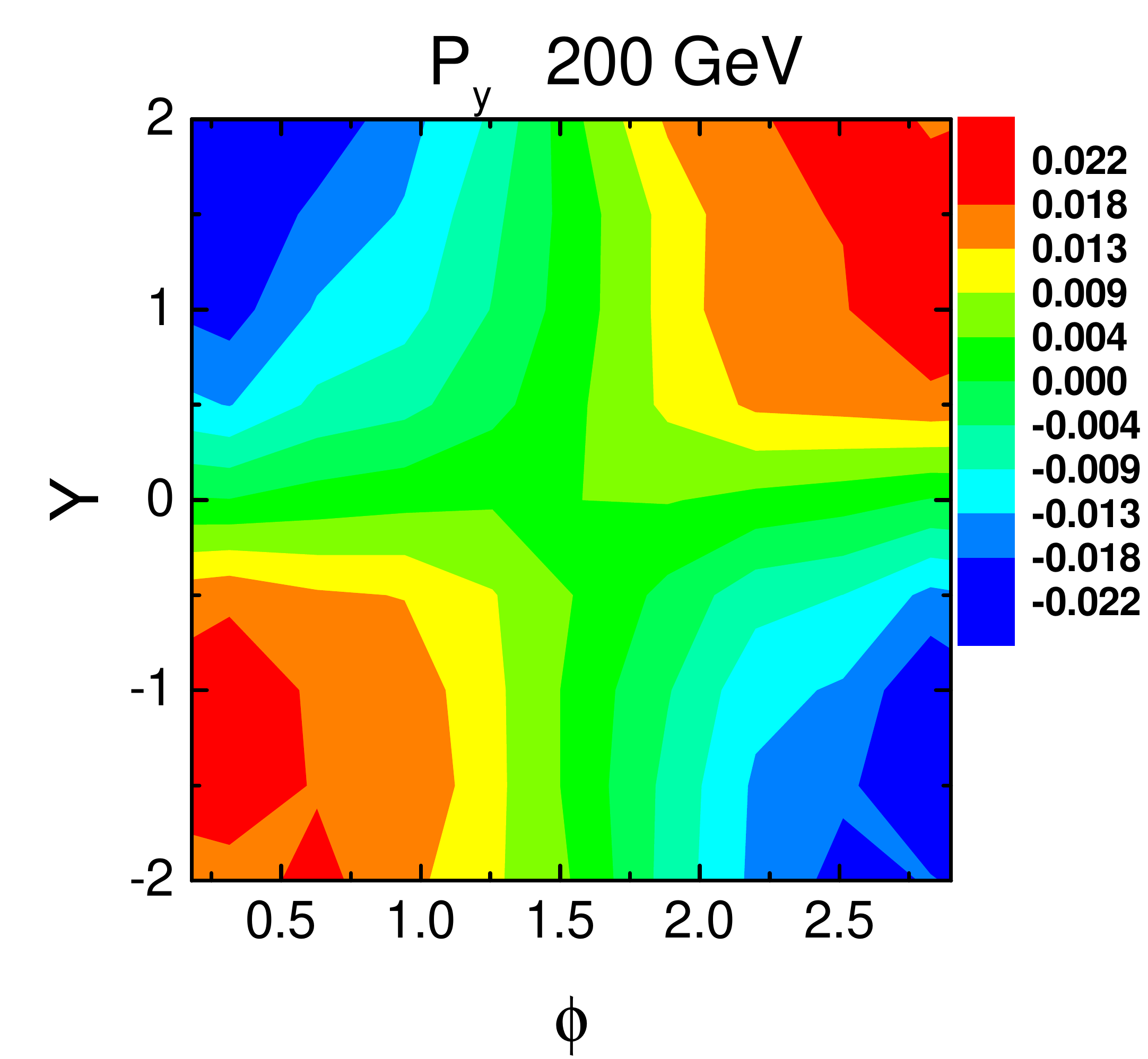}$\;\;\;\;$
\includegraphics[height=4.6cm]{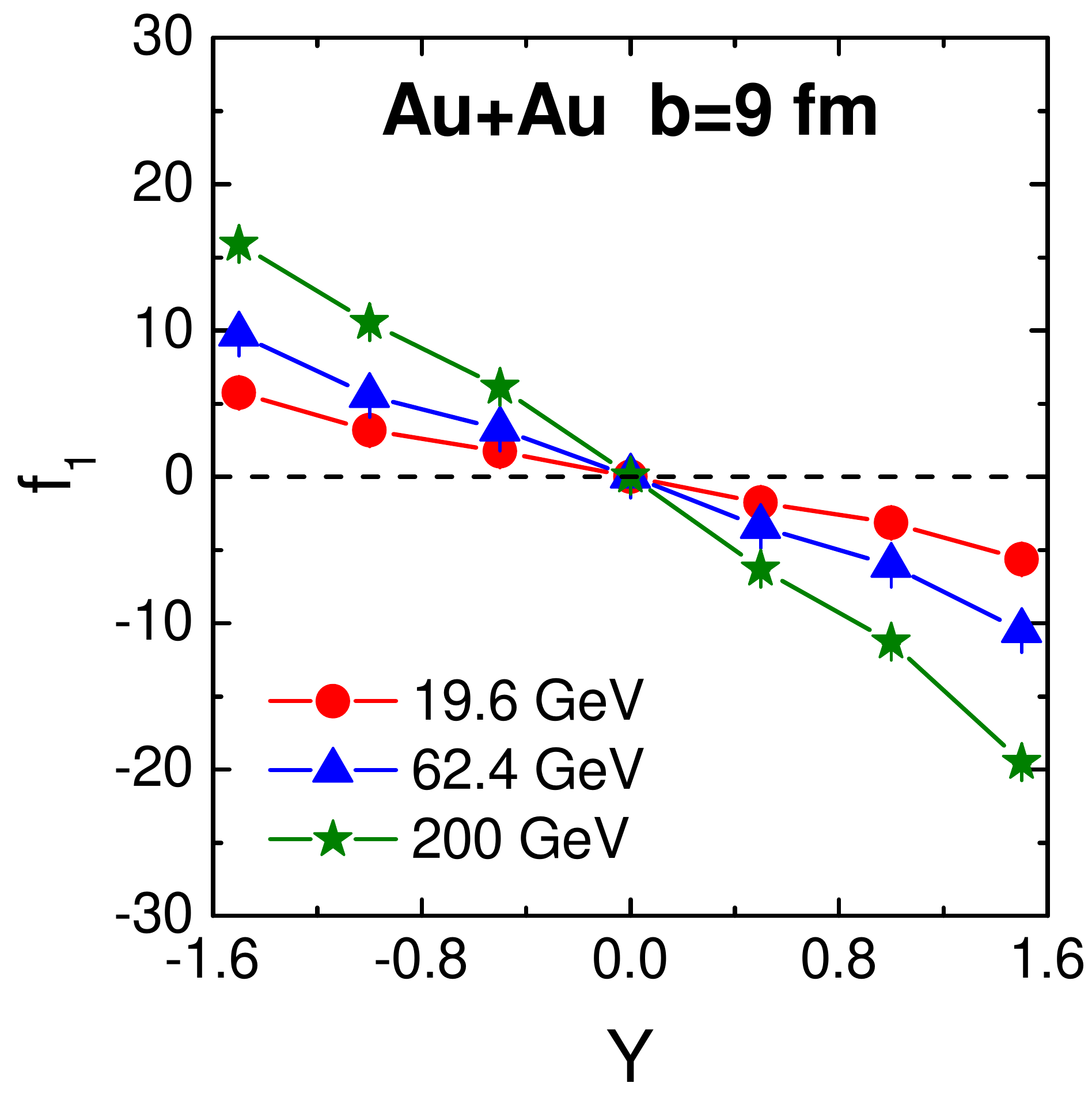}$\;\;\;\;$
\caption{Left: the polarization $P_y$ on $\phi-Y$ plane which is the spin-polarization response to the upper-right panel of \fig{th-spa-eta} up to the linear order in thermal vorticity according to \eq{spin}. Right: the directed spin flow $f_1$ defined in \eq{harmo} versus rapidity $Y$ \cite{Wei:2018zfb}.}
\label{pol:local}
\end{center}
\end{figure}

\fig{pol:local} shows another way to present the local polarization \cite{Wei:2018zfb}. In the left panel, we present the distribution of the transverse spin polarization $P_y$ on the $\phi-Y$ plane where $\phi$ is the momentum azimuthal angle and $Y$ is the rapidity. Clearly, as a spin-polarization response to the quadrupole structure of the vorticity field shown in the upper-right panel of \fig{th-spa-eta}, $P_y(Y,\phi)$ also shows a quadrupole structure. To characterize such a nontrivial $\phi$ dependence of $P_y(\phi)$ at a given rapidity $Y$, we can decompose $P_y(Y,\phi)$ into a harmonic series,
\begin{eqnarray}
\label{harmo}
P_{y}(Y,\phi)= \frac{1}{2\pi}\frac{dP_{y}}{dY}\{1+2\sum_{n=1}^{\infty}f_{n}\cos[n(\phi-\Phi_{n})]\},
\end{eqnarray}
where $\Phi_n$ defines the $n$-th harmonic plane for spin with the corresponding harmonic coefficient $f_n$. The first harmonic coefficient, $f_1$, shown in the right panel of \fig{pol:local}, is induced by the vorticity from collective expansion, which is odd in rapidity and peaks at finite rapidity in accordance with \fig{th-spa-eta}. The measurement of such a directed flow of spin polarization would be the indicator of the quadrupole structure in the vorticity field due to inhomogeneous expansion of the fireball.

\begin{figure}[!htb]
\begin{center}
\includegraphics[height=5cm]{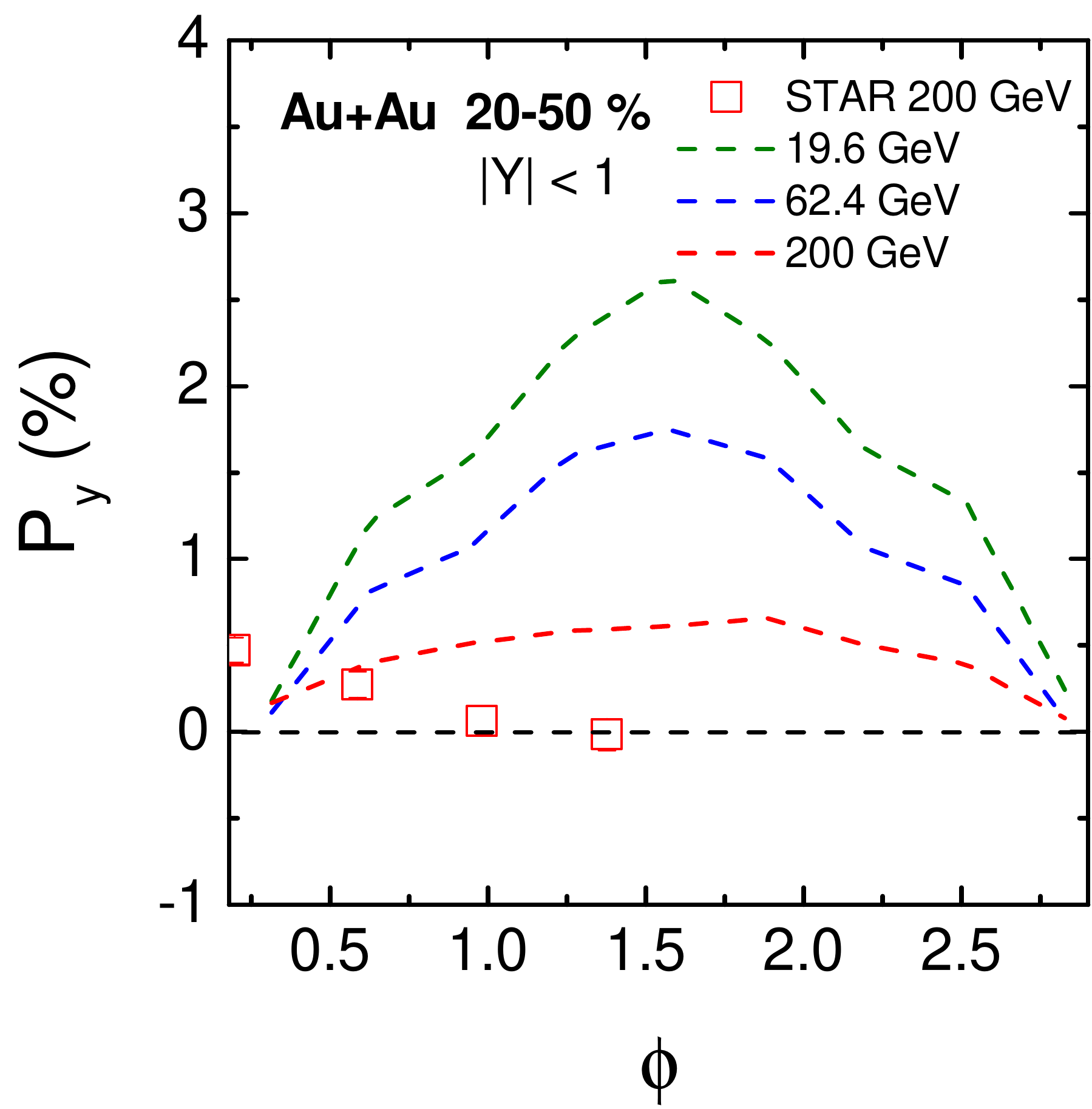}$\;\;\;\;$
\includegraphics[height=5cm]{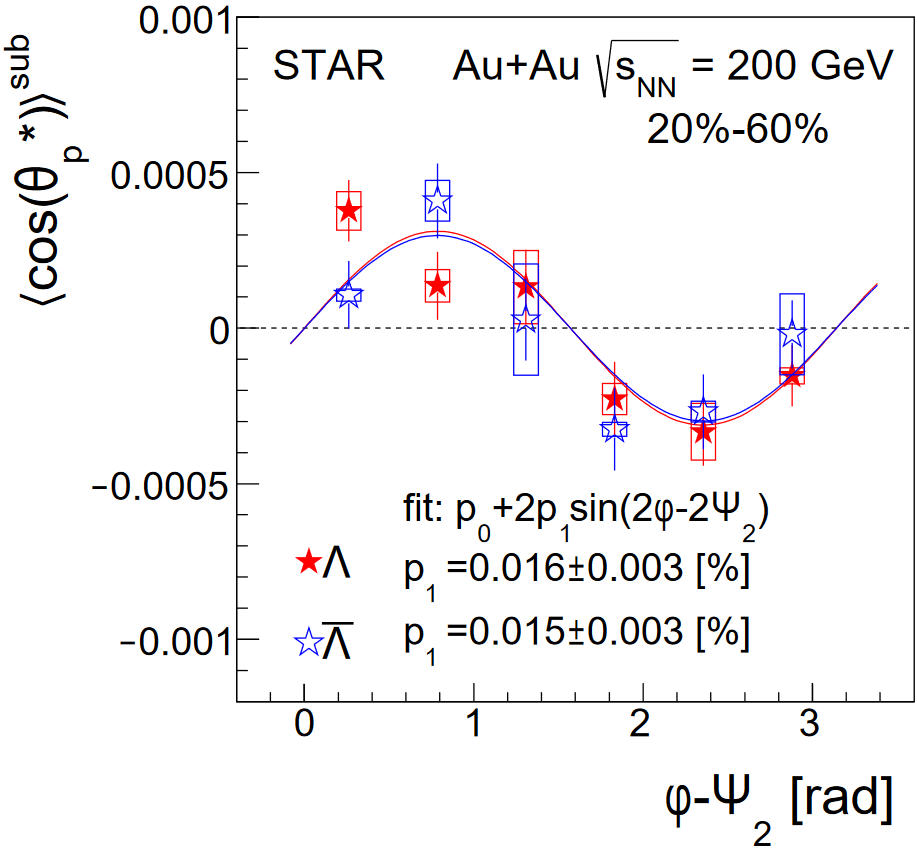}$\;\;\;\;$
\caption{Left: the polarization $P_y$ as a function of the azimuthal angle $\phi$. The red squares are experimental data~\cite{Adam:2018ivw}. Right: the experimental results of longitudinal polarization $P_z(\phi)$ from STAR Collaboration~\cite{Adam:2019srw}. Note that $P_z\sim \langle\cos\theta_p^*\rangle$.}
\label{pol:local2}
\end{center}
\end{figure}

\textit{The ``sign problem".} Although the global polarization $P_y$ can be described well by simulations based on the thermal vorticity following Eq. (\ref{average-spin-vector}), such relation fails in describing the azimuthal dependence of $P_y$ at mid-rapidity. In fact, the theoretical calculations, including both the transport-model and hydrodynamic model calculations, found that $P_y(\phi)$ at mid-rapidity grows from $\phi=0$ (i.e., the in-plane direction) to $\phi=\pi/2$ (i.e., the out-of-plane direction) while the experimental data shows an opposite trend, see the left panel of \fig{pol:local2}. In addition, the longitudinal polarization $P_z(\phi)$ at mid-rapidity also has a similar ``sign problem": the theoretical calculations predicted that $P_z(\phi)\sim -\sin (2\phi )$ as shown in Fig. \ref{pol:local-Xia}. This can also be seen in \fig{th-spa-long} as the spin polarization is roughly proportional to the thermal vorticity following Eq. (\ref{average-spin-vector}). However the data show an opposite sign, see the right panel of \fig{pol:local2}. These sign problems challenge the thermal vorticity interpretation of the measured $\Lambda$ polarization and are puzzles at the moment.

Here we have several comments about them. (1) As we have already discussed, the feed-down decays of other strange baryons constitute a major contribution to the total yield of $\Lambda$ and $\bar\Lambda$. Thus, to bridge the measured spin polarization and the vorticity, we must take into account the feed-down contributions. In addition, $\Lambda$ hyperon produced from the feed-down decay can have opposite spin polarization comparing to its parent particle in some decay channels, e.g., $\Sigma^0\rightarrow\Lambda+\gamma$. Recently, the feed-down effects have been carefully studied in Refs.~\cite{Xia:2019fjf,Becattini:2019ntv}. Although the feed-down contribution suppresses the polarization of primordial $\Lambda$, it is not strong enough to resolve the sign problem. (2) At the moment, most of the theoretical studies are based on Eq. (\ref{average-spin-vector}) which assumes global equilibrium for the spin degree of freedom. This might not be the case for realistic heavy ion collisions. In non-equilibrium state or even near equilibrium state, the spin polarization is not determined by the thermal vorticity and should be treated as an independent dynamic variable. This requires new theoretical frameworks like the spin hydrodynamics; see e.g.~\cite{Florkowski:2017ruc,Hattori:2019lfp}. Recently, there have been progresses in developing these new frameworks and hopefully the numerical simulations based on them can give more accurate description of the $\Lambda$ polarization and insight to the sign problem. (3) There have been theoretical explanations of the sign problem based on chiral kinetic theory~\cite{Sun:2018bjl,Liu:2019krs}, blast-wave model~\cite{Adam:2019srw} and hydrodynamics~\cite{Florkowski:2019voj,Wu:2019eyi}. But they introduce new assumptions such as the presence of net chirality~\cite{Sun:2018bjl}, kinematic vorticity or T-vorticity dominance of the polarization~\cite{Adam:2019srw,Florkowski:2019voj,Wu:2019eyi}) which need further examinations; see Refs.~\cite{Wang:2017jpl,Liu:2020ymh,Huang:2020xyr,Becattini:2020ngo,Gao:2020vbh} for recent reviews.


\textit{The magnetic polarization.} Finally let us discuss one intriguing aspect of the measured global polarization: there is a visible difference between hyperons and anti-hyperons, especially in the low beam energy region. While  the error bars are still  too large to unambiguously identify a splitting between $P_{\bar{\Lambda}}$ and $P_\Lambda$, the difference shown by these data is significant enough to warrant a serious investigation into the probable causes.  One natural and plausible explanation could be the magnetic polarization effect (which distinguishes particles from anti-particles) in addition to the rotational polarization (which is ``blind'' to particle/anti-particle identities)~\cite{Becattini:2016gvu,Muller:2018ibh,Guo:2019joy,Guo:2019mgh}. Indeed the hyperon $\Lambda$ and anti-hyperon $\bar{\Lambda}$ have negative and positive magnetic moments respectively. When subject to an external magnetic field, $\bar{\Lambda}$ spin would be more aligned along the field direction while $\bar{\Lambda}$ spin would be more aligned  against the field direction. This could indeed qualitatively explain the observed splitting with $P_{\bar{\Lambda}} > P_\Lambda$, provided that the magnetic field in heavy ion collisions is indeed approximately parallel to average vorticity and possibly survive long enough till the freeze-out time.

\begin{figure}[!htb]
\begin{center}
\includegraphics[height=5cm]{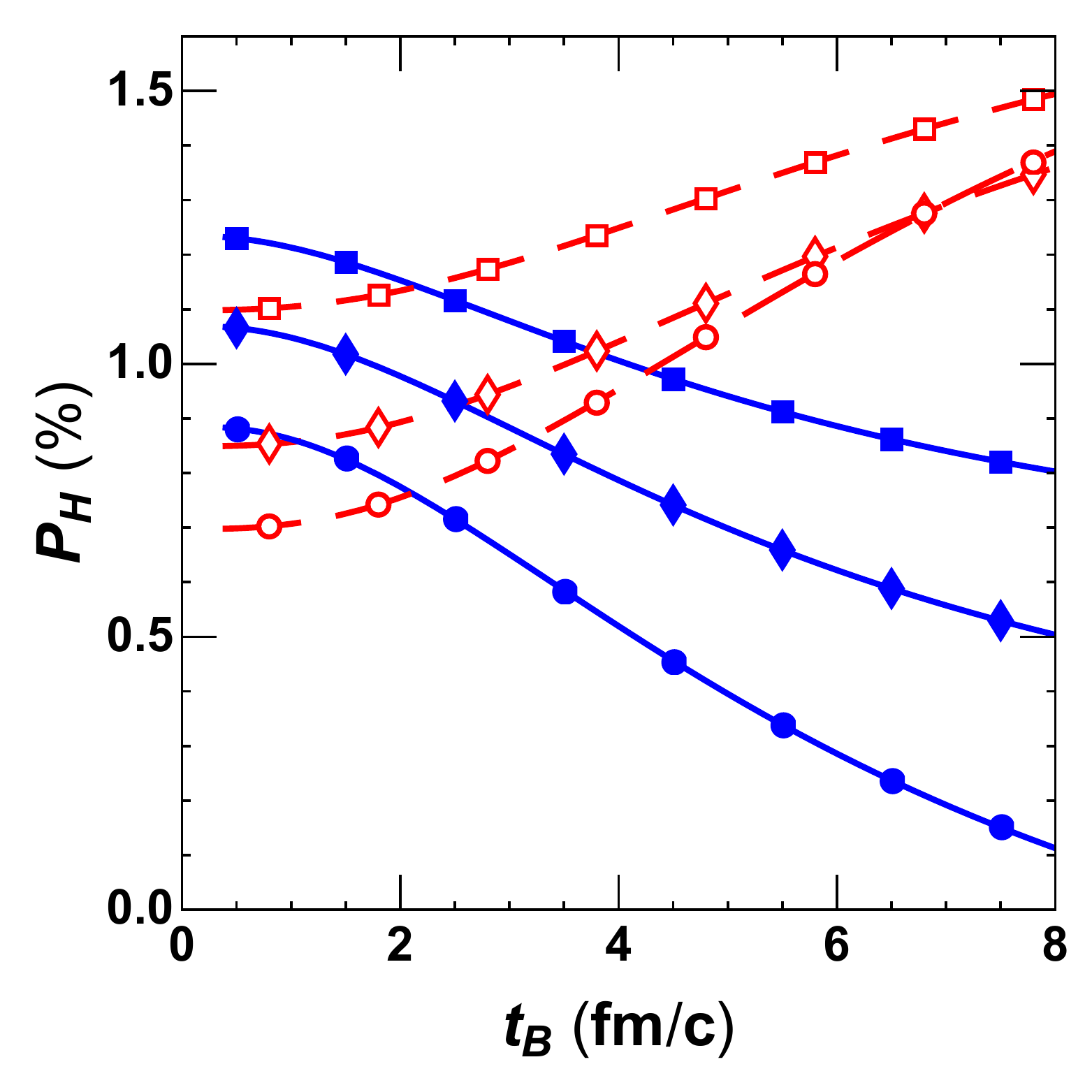}$\;\;\;\;$
\includegraphics[height=5cm]{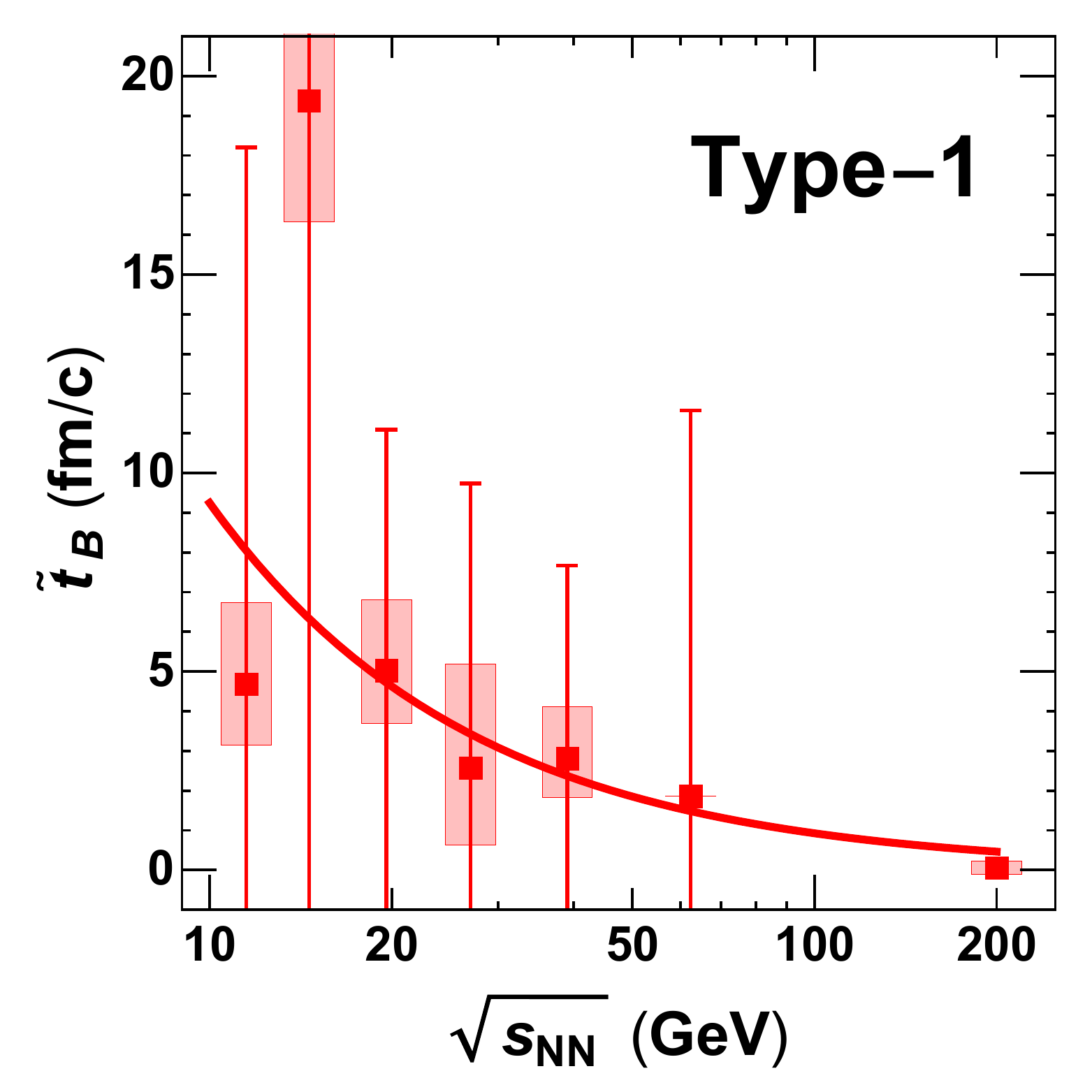}$\;\;\;\;$
\caption{Left: The dependence on magnetic field lifetime parameter $t_B$ of the global polarization signals $P_H$ for hyperons ( $H \to \Lambda$, blue solid curves with filled symbols) and anti-hyperons ( $H \to \bar{\Lambda}$, red dashed curves with open symbols)  at  beam energy
$\sqrt{s_{NN}}=$19.6 (square), 27 (diamond), 39 (circle) GeV respectively~\cite{Guo:2019joy}.
Right:
The optimal value of magnetic field lifetime parameter $\tilde{t}_B$ extracted from polarization splitting $\Delta P$ data for a range of collision beam energy $\sqrt{s_{NN}}$. The results in this plot use the parameterization $eB(t) = eB(0) / [1+(t-t_0)^2/t_B^2]$ ( --- see \cite{Guo:2019joy} for details).  The solid curve is from fitting analysis with a formula $\tilde{t}_B = \frac{A}{\sqrt{s_{NN}}}$. The error bars are converted from the corresponding errors of experimental data in \cite{STAR:2017ckg,Adam:2018ivw}.}
\label{pol:mag}
\end{center}
\end{figure}

To examine whether this idea may work,  quantitative simulations have been carried out recently within the AMPT framework~\cite{Guo:2019joy}. Under the presence of both vorticity and magnetic field, the polarization given in  Eq.~(\ref{spin}) should be modified to include both effects as follows:
\begin{eqnarray}
\label{spin-mag}
S^\mu(x,p)=-\frac{s(s+1)}{6m}(1-n_F)\epsilon^{\mu\nu\rho\sigma} p_\nu \left [ \varpi_{\rho\sigma}(x)  \mp 2 (eF_{\rho\sigma})\  \mu_\Lambda/T_f \right ] \ ,
\end{eqnarray}
where the $\mp$ sign is for $\Lambda$ and $\bar{\Lambda}$ while $\mu_\Lambda = 0.613/(2m_N)$ is the absolute value of the hyperon/anti-hyperon magnetic moment, with $m_N=938\rm MeV$ being the nucleon mass.  $T_f$ is the local temperature upon the particle's formation. Here we focus on the electromagnetic field component that is most relevant to the global polarization effect, namely $eB_y = eF_{31} = - eF_{13}$ along the out-of-plane direction. By adopting a certain parameterization of the time dependence for the magnetic field with a lifetime parameter $t_B$, one could then investigation how the polarization splitting depends on the $B$ field lifetime.  The left plan of  Fig.~\ref{pol:mag}  shows how  the magnetic field lifetime $t_B$ would quantitatively  influence the polarization of $\Lambda$ and $\bar{\Lambda}$.
As one can see, with increasing magnetic field lifetime (which means stronger magnetic field at late time in the collisions), the $P_{\bar{\Lambda}}$ steadily increases while the $P_\Lambda$ decreases at all collision energies. With long enough $t_B$, eventually the $P_{\bar{\Lambda}}$ always becomes larger than $P_\Lambda$. By comparing with experimentally measured polarization splitting, one could actually extract the optimal value (or a constraint) on the magnetic field lifetime. Such analysis is shown in the right panel of  Fig.~\ref{pol:mag}.  A number of different parameterizations were studied in \cite{Guo:2019joy} and the overall analysis suggests an empirical formula for possible magnetic field lifetime:  $t_B = A/\sqrt{s_{NN}}$ with $A=115\pm 16\ \rm GeV\cdot fm/c$.  Interestingly, this is considerably longer than the expected vacuum magnetic field lifetime without any medium effect, which could be estimated by $t_{vac}\simeq 2R_A/\gamma \simeq 26\ {\rm GeV\cdot fm\cdot c^{-1}}/\sqrt{s_{NN}}$. Such extended magnetic field lifetime, as indicated by polarization difference, may imply a considerable role of the medium-generated   dynamical magnetic field especially at low beam energy~\cite{Guo:2019mgh}.

\section{Summary}
\label{sec:sum}
The non-vanishing global spin polarization of $\Lambda$ and $\bar{\Lambda}$ has been
measured by STAR collaboration in Au+Au collisions at $\sqrt{s_{NN}}=7.7-200$
GeV. Microscopically, such a global polarization originates from the spin-orbit coupling of particle scatterings
in a fluid with local vorticity. It has been shown that the spin-orbit coupling can lead to
a spin-vorticity coupling when taking an ensemble average over random
incoming momenta of scattering particles in a locally thermalized fluid.
With the spin-vorticity coupling, the local spin polarization can be obtained from
the vorticity field in the fluid. The global polarization is an integration of the local one
in whole phase space. In this note we review recent progress on the vorticity formation
and spin polarization in heavy ion collisions
with transport models. We present an introduction of the fluid vorticity
in non-relativistic and relativistic hydrodynamics.
We discuss  the spin polarization in a vortical fluid in the Wigner function formalism
for massive spin-1/2 fermions, in which we derive the freeze-out formula for the spin polarization
in heavy ion collisions. Then we show results for various properties of  the kinematic and thermal vorticity with transport models,
including: the evolution in time and space, the correlation to the participant plane,
the collision energy dependence, etc.
Finally we give a brief overview of recent theoretical results for the spin polarization
of $\Lambda$ and $\bar{\Lambda}$ including the global and local polarization,
the polarization harmonics in azimuthal angles, the sign problem in the longitudinal polarization as well as the polarization difference between particles and anti-particles.

\section*{Acknowledgements}
We thank Wei-Tian Deng, Hui Li, Yu-Chen Liu, Yin Jiang, Zi-Wei Lin, Long-Gang Pang, Shuzhe Shi, De-Xian Wei, Xin-Nian Wang for collaborations and discussions. This work is supported in part by the NSFC Grants No.~11535012, No.~11675041 and No. 11890713, as well as by the NSF Grant No. PHY-1913729 and by the U.S. Department of Energy, Office of Science, Office of Nuclear Physics, within the framework of the Beam Energy Scan Theory (BEST) Topical Collaboration.



\begin{thebibliography}{999}%

\bibitem{Liang:2004ph}
Z.~T.~Liang and X.~N.~Wang,
Phys. Rev. Lett. \textbf{94}, 102301 (2005)
[arXiv:nucl-th/0410079 [nucl-th]].

\bibitem{Liang:2004xn}
Z.~T.~Liang and X.~N.~Wang,
Phys. Lett. B \textbf{629}, 20-26 (2005)
[arXiv:nucl-th/0411101 [nucl-th]].

\bibitem{Voloshin:2004ha}
S.~A.~Voloshin,
[arXiv:nucl-th/0410089 [nucl-th]].

\bibitem{Betz:2007kg}
B.~Betz, M.~Gyulassy and G.~Torrieri,
Phys. Rev. C \textbf{76}, 044901 (2007)
[arXiv:0708.0035 [nucl-th]].

\bibitem{Becattini:2007sr}
F.~Becattini, F.~Piccinini and J.~Rizzo,
Phys. Rev. C \textbf{77}, 024906 (2008)
[arXiv:0711.1253 [nucl-th]].

\bibitem{Gao:2007bc}
J.~H.~Gao, S.~W.~Chen, W.~t.~Deng, Z.~T.~Liang, Q.~Wang and X.~N.~Wang,
Phys. Rev. C \textbf{77}, 044902 (2008)
[arXiv:0710.2943 [nucl-th]].

\bibitem{Huang:2011ru}
X.~G.~Huang, P.~Huovinen and X.~N.~Wang,
Phys. Rev. C \textbf{84}, 054910 (2011)
[arXiv:1108.5649 [nucl-th]].

\bibitem{Chen:2008wh}
S.~w.~Chen, J.~Deng, J.~h.~Gao and Q.~Wang,
Front. Phys. China \textbf{4}, 509-516 (2009)
[arXiv:0801.2296 [hep-ph]].

\bibitem{STAR:2017ckg}
L.~Adamczyk \textit{et al.} [STAR],
Nature \textbf{548}, 62-65 (2017)
[arXiv:1701.06657 [nucl-ex]].

\bibitem{Adam:2018ivw}
J.~Adam \textit{et al.} [STAR],
Phys. Rev. C \textbf{98}, 014910 (2018)
[arXiv:1805.04400 [nucl-ex]].

\bibitem{Acharya:2019ryw}
S.~Acharya \textit{et al.} [ALICE],
Phys. Rev. C \textbf{101}, 044611 (2020)
[arXiv:1909.01281 [nucl-ex]].

\bibitem{Zhang:2019xya}
J.~j.~Zhang, R.~h.~Fang, Q.~Wang and X.~N.~Wang,
Phys. Rev. C \textbf{100}, 064904 (2019)
[arXiv:1904.09152 [nucl-th]].

\bibitem{Baznat:2013zx}
M.~Baznat, K.~Gudima, A.~Sorin and O.~Teryaev,
Phys. Rev. C \textbf{88}, 061901 (2013)
[arXiv:1301.7003 [nucl-th]].

\bibitem{Csernai:2013bqa}
L.~P.~Csernai, V.~K.~Magas and D.~J.~Wang,
Phys. Rev. C \textbf{87}, 034906 (2013)
[arXiv:1302.5310 [nucl-th]].

\bibitem{Csernai:2014ywa}
L.~P.~Csernai, D.~J.~Wang, M.~Bleicher and H.~StÃ¶cker,
Phys. Rev. C \textbf{90}, 021904 (2014).

\bibitem{Becattini:2015ska}
F.~Becattini, G.~Inghirami, V.~Rolando, A.~Beraudo, L.~Del Zanna, A.~De Pace, M.~Nardi, G.~Pagliara and V.~Chandra,
Eur. Phys. J. C \textbf{75}, 406 (2015)
[arXiv:1501.04468 [nucl-th]].

\bibitem{Teryaev:2015gxa}
O.~Teryaev and R.~Usubov,
Phys. Rev. C \textbf{92}, 014906 (2015).

\bibitem{Jiang:2016woz}
Y.~Jiang, Z.~W.~Lin and J.~Liao,
Phys. Rev. C \textbf{94}, 044910 (2016)
[arXiv:1602.06580 [hep-ph]].

\bibitem{Deng:2016gyh}
W.~T.~Deng and X.~G.~Huang,
Phys. Rev. C \textbf{93}, 064907 (2016)
[arXiv:1603.06117 [nucl-th]].

\bibitem{Ivanov:2017dff}
Y.~B.~Ivanov and A.~A.~Soldatov,
Phys. Rev. C \textbf{95}, 054915 (2017)
[arXiv:1701.01319 [nucl-th]].

\bibitem{Deng:2020ygd}
X.~G.~Deng, X.~G.~Huang, Y.~G.~Ma and S.~Zhang,
Phys. Rev. C \textbf{101}, 064908 (2020)
[arXiv:2001.01371 [nucl-th]].

\bibitem{Li:2017slc}
H.~Li, L.~G.~Pang, Q.~Wang and X.~L.~Xia,
Phys. Rev. C \textbf{96}, 054908 (2017)
[arXiv:1704.01507 [nucl-th]].

\bibitem{Wei:2018zfb}
D.~X.~Wei, W.~T.~Deng and X.~G.~Huang,
Phys. Rev. C \textbf{99}, 014905 (2019)
[arXiv:1810.00151 [nucl-th]].

\bibitem{Shi:2017wpk}
S.~Shi, K.~Li and J.~Liao,
Phys. Lett. B \textbf{788}, 409-413 (2019)
[arXiv:1712.00878 [nucl-th]].


\bibitem{Karpenko:2016jyx}
I.~Karpenko and F.~Becattini,
Eur. Phys. J. C \textbf{77}, 213 (2017)
[arXiv:1610.04717 [nucl-th]].

\bibitem{Xie:2017upb}
Y.~Xie, D.~Wang and L.~P.~Csernai,
Phys. Rev. C \textbf{95}, 031901 (2017)
[arXiv:1703.03770 [nucl-th]].

\bibitem{Sun:2017xhx}
Y.~Sun and C.~M.~Ko,
Phys. Rev. C \textbf{96}, 024906 (2017)
[arXiv:1706.09467 [nucl-th]].

\bibitem{Xie:2016fjj}
Y.~L.~Xie, M.~Bleicher, H.~St\"ocker, D.~J.~Wang and L.~P.~Csernai,
Phys. Rev. C \textbf{94}, 054907 (2016)
[arXiv:1610.08678 [nucl-th]].

\bibitem{Moffatt:1969}
H.~K.~Moffatt, J. Fluid Mech. {\bf 35}, 117 (1969).

\bibitem{Moreau:1961}
J.~J.~Moreau, C. R. Acad. Sci. Paris, {\bf 252}, 2810 (1961).

\bibitem{Heinz:1983nx}
U.~W.~Heinz,
Phys. Rev. Lett. \textbf{51}, 351 (1983).

\bibitem{Elze:1986qd}
H.~T.~Elze, M.~Gyulassy and D.~Vasak,
Nucl. Phys. B \textbf{276}, 706-728 (1986).

\bibitem{Vasak:1987um}
D.~Vasak, M.~Gyulassy and H.~T.~Elze,
Annals Phys. \textbf{173}, 462-492 (1987).

\bibitem{Zhuang:1995pd}
P.~Zhuang and U.~W.~Heinz,
Annals Phys. \textbf{245}, 311-338 (1996)
[arXiv:nucl-th/9502034 [nucl-th]].

\bibitem{Gao:2012ix}
J.~H.~Gao, Z.~T.~Liang, S.~Pu, Q.~Wang and X.~N.~Wang,
Phys. Rev. Lett. \textbf{109}, 232301 (2012)
[arXiv:1203.0725 [hep-ph]].

\bibitem{Chen:2012ca}
J.~W.~Chen, S.~Pu, Q.~Wang and X.~N.~Wang,
Phys. Rev. Lett. \textbf{110}, 262301 (2013)
[arXiv:1210.8312 [hep-th]].

\bibitem{Gao:2015zka}
J.~h.~Gao and Q.~Wang,
Phys. Lett. B \textbf{749}, 542-546 (2015)
[arXiv:1504.07334 [nucl-th]].

\bibitem{Fang:2016vpj}
R.~h.~Fang, L.~g.~Pang, Q.~Wang and X.~n.~Wang,
Phys. Rev. C \textbf{94}, 024904 (2016)
[arXiv:1604.04036 [nucl-th]].

\bibitem{Hidaka:2016yjf}
Y.~Hidaka, S.~Pu and D.~L.~Yang,
Phys. Rev. D \textbf{95}, 091901 (2017)
[arXiv:1612.04630 [hep-th]].

\bibitem{Mueller:2017lzw}
N.~Mueller and R.~Venugopalan,
Phys. Rev. D \textbf{97}, 051901 (2018)
[arXiv:1701.03331 [hep-ph]].

\bibitem{Huang:2018wdl}
A.~Huang, S.~Shi, Y.~Jiang, J.~Liao and P.~Zhuang,
Phys. Rev. D \textbf{98}, 036010 (2018)
[arXiv:1801.03640 [hep-th]].

\bibitem{Liu:2018xip}
Y.~C.~Liu, L.~L.~Gao, K.~Mameda and X.~G.~Huang,
Phys. Rev. D \textbf{99}, 085014 (2019)
[arXiv:1812.10127 [hep-th]].

\bibitem{Gao:2017gfq}
J.~h.~Gao, S.~Pu and Q.~Wang,
Phys. Rev. D \textbf{96}, 016002 (2017)
[arXiv:1704.00244 [nucl-th]].

\bibitem{Gao:2018wmr}
J.~H.~Gao, Z.~T.~Liang, Q.~Wang and X.~N.~Wang,
Phys. Rev. D \textbf{98}, 036019 (2018)
[arXiv:1802.06216 [hep-ph]].


\bibitem{Weickgenannt:2019dks}
N.~Weickgenannt, X.~L.~Sheng, E.~Speranza, Q.~Wang and D.~H.~Rischke,
Phys. Rev. D \textbf{100}, 056018 (2019)
[arXiv:1902.06513 [hep-ph]].

\bibitem{Gao:2019znl}
J.~H.~Gao and Z.~T.~Liang,
Phys. Rev. D \textbf{100}, 056021 (2019)
[arXiv:1902.06510 [hep-ph]].

\bibitem{Hattori:2019ahi}
K.~Hattori, Y.~Hidaka and D.~L.~Yang,
Phys. Rev. D \textbf{100}, 096011 (2019)
[arXiv:1903.01653 [hep-ph]].

\bibitem{Wang:2019moi}
Z.~Wang, X.~Guo, S.~Shi and P.~Zhuang,
Phys. Rev. D \textbf{100}, 014015 (2019)
[arXiv:1903.03461 [hep-ph]].

\bibitem{Liu:2020flb}
Y.~C.~Liu, K.~Mameda and X.~G.~Huang,
Chin. Phys. C \textbf{44}, 094101 (2020)
[arXiv:2002.03753 [hep-ph]].

\bibitem{Sheng:2020oqs}
X.~L.~Sheng, Q.~Wang and X.~G.~Huang,
Phys. Rev. D \textbf{102}, no.2, 025019 (2020)
doi:10.1103/PhysRevD.102.025019
[arXiv:2005.00204 [hep-ph]].


\bibitem{Florkowski:2018ahw}
W.~Florkowski, A.~Kumar and R.~Ryblewski,
Phys. Rev. C \textbf{98}, no.4, 044906 (2018)
doi:10.1103/PhysRevC.98.044906
[arXiv:1806.02616 [hep-ph]].

\bibitem{Yang:2020hri}
D.~L.~Yang, K.~Hattori and Y.~Hidaka,
JHEP \textbf{20}, 070 (2020)
doi:10.1007/JHEP07(2020)070
[arXiv:2002.02612 [hep-ph]].

\bibitem{Weickgenannt:2020aaf}
N.~Weickgenannt, E.~Speranza, X.~l.~Sheng, Q.~Wang and D.~H.~Rischke,
[arXiv:2005.01506 [hep-ph]].

\bibitem{Wang:2020pej}
Z.~Wang, X.~Guo and P.~Zhuang,
[arXiv:2009.10930 [hep-th]].

\bibitem{Becattini:2013fla}
F.~Becattini, V.~Chandra, L.~Del Zanna and E.~Grossi,
Annals Phys. \textbf{338}, 32-49 (2013)
[arXiv:1303.3431 [nucl-th]].

\bibitem{Pang:2016igs}
L.~G.~Pang, H.~Petersen, Q.~Wang and X.~N.~Wang,
Phys. Rev. Lett. \textbf{117}, 192301 (2016)
[arXiv:1605.04024 [hep-ph]].

\bibitem{Becattini:2017gcx}
F.~Becattini and I.~Karpenko,
Phys. Rev. Lett. \textbf{120}, 012302 (2018)
[arXiv:1707.07984 [nucl-th]].

\bibitem{Xia:2018tes}
X.~L.~Xia, H.~Li, Z.~B.~Tang and Q.~Wang,
Phys. Rev. C \textbf{98}, 024905 (2018)
[arXiv:1803.00867 [nucl-th]].

\bibitem{DelZanna:2013eua}
L.~Del Zanna, V.~Chandra, G.~Inghirami, V.~Rolando, A.~Beraudo, A.~De Pace, G.~Pagliara, A.~Drago and F.~Becattini,
Eur. Phys. J. C \textbf{73}, 2524 (2013)
[arXiv:1305.7052 [nucl-th]].

\bibitem{Pang:2012he}
L.~Pang, Q.~Wang and X.~N.~Wang,
Phys. Rev. C \textbf{86}, 024911 (2012)
[arXiv:1205.5019 [nucl-th]].

\bibitem{Pang:2018zzo}
L.~G.~Pang, H.~Petersen and X.~N.~Wang,
Phys. Rev. C \textbf{97}, 064918 (2018)
[arXiv:1802.04449 [nucl-th]].

\bibitem{Hirano:2012kj}
T.~Hirano, P.~Huovinen, K.~Murase and Y.~Nara,
Prog. Part. Nucl. Phys. \textbf{70}, 108-158 (2013)
[arXiv:1204.5814 [nucl-th]].


\bibitem{Bloczynski:2012en}
J.~Bloczynski, X.~G.~Huang, X.~Zhang and J.~Liao,
Phys. Lett. B \textbf{718}, 1529-1535 (2013)
[arXiv:1209.6594 [nucl-th]].

\bibitem{Bloczynski:2013mca}
J.~Bloczynski, X.~G.~Huang, X.~Zhang and J.~Liao,
Nucl. Phys. A \textbf{939}, 85-100 (2015)
[arXiv:1311.5451 [nucl-th]].

\bibitem{Deng:2012pc}
W.~T.~Deng and X.~G.~Huang,
Phys. Rev. C \textbf{85}, 044907 (2012)
[arXiv:1201.5108 [nucl-th]].

\bibitem{Abelev:2007zk}
B.~I.~Abelev \textit{et al.} [STAR],
Phys. Rev. C \textbf{76}, 024915 (2007)
[arXiv:0705.1691 [nucl-ex]].

\bibitem{Siddique:2017ddr}
I.~Siddique, Z.~T.~Liang, M.~A.~Lisa, Q.~Wang and Z.~B.~Xu,
Chin. Phys. C \textbf{43}, 014103 (2019)
[arXiv:1710.00134 [nucl-th]].

\bibitem{Sheng:2019kmk}
X.~L.~Sheng, L.~Oliva and Q.~Wang,
Phys. Rev. D \textbf{101}, 096005 (2020)
[arXiv:1910.13684 [nucl-th]].

\bibitem{Sheng:2020ghv}
X.~L.~Sheng, Q.~Wang and X.~N.~Wang,
Phys. Rev. D \textbf{102}, 056013 (2020)
[arXiv:2007.05106 [nucl-th]].

\bibitem{Xia:2020tyd}
X.~L.~Xia, H.~Li, X.~G.~Huang and H.~Z.~Huang,
[arXiv:2010.01474 [nucl-th]].

\bibitem{Taya:2020sej}
H.~Taya \textit{et al.} [ExHIC-P],
Phys. Rev. C \textbf{102}, 021901 (2020)
[arXiv:2002.10082 [nucl-th]].

\bibitem{Ipp:2007ng}
A.~Ipp, A.~Di Piazza, J.~Evers and C.~H.~Keitel,
Phys. Lett. B \textbf{666}, 315-319 (2008)
[arXiv:0710.5700 [hep-ph]].

\bibitem{Becattini:2016gvu}
F.~Becattini, I.~Karpenko, M.~Lisa, I.~Upsal and S.~Voloshin,
Phys. Rev. C \textbf{95}, 054902 (2017)
[arXiv:1610.02506 [nucl-th]].


\bibitem{Xia:2019fjf}
X.~L.~Xia, H.~Li, X.~G.~Huang and H.~Z.~Huang,
Phys. Rev. C \textbf{100}, 014913 (2019)
[arXiv:1905.03120 [nucl-th]].

\bibitem{Becattini:2019ntv}
F.~Becattini, G.~Cao and E.~Speranza,
Eur. Phys. J. C \textbf{79}, 741 (2019)
[arXiv:1905.03123 [nucl-th]].


\bibitem{HADES:2019}
	F.~Kornas for HADES Collaboration, Talk given at {\it Strange Quark Matter 2019}, Bali, Italy, June 11-15, 2019.

\bibitem{Florkowski:2017ruc}
W.~Florkowski, B.~Friman, A.~Jaiswal and E.~Speranza,
Phys. Rev. C \textbf{97}, 041901 (2018)
[arXiv:1705.00587 [nucl-th]].

\bibitem{Hattori:2019lfp}
K.~Hattori, M.~Hongo, X.~G.~Huang, M.~Matsuo and H.~Taya,
Phys. Lett. B \textbf{795}, 100-106 (2019)
[arXiv:1901.06615 [hep-th]].

\bibitem{Sun:2018bjl}
Y.~Sun and C.~M.~Ko,
Phys. Rev. C \textbf{99}, 011903 (2019)
[arXiv:1810.10359 [nucl-th]].


\bibitem{Liu:2019krs}
S.~Y.~F.~Liu, Y.~Sun and C.~M.~Ko,
Phys. Rev. Lett. \textbf{125}, 062301 (2020)
[arXiv:1910.06774 [nucl-th]].

\bibitem{Adam:2019srw}
J.~Adam \textit{et al.} [STAR],
Phys. Rev. Lett. \textbf{123}, 132301 (2019)
[arXiv:1905.11917 [nucl-ex]].

\bibitem{Florkowski:2019voj}
W.~Florkowski, A.~Kumar, R.~Ryblewski and A.~Mazeliauskas,
Phys. Rev. C \textbf{100}, 054907 (2019)
[arXiv:1904.00002 [nucl-th]].

\bibitem{Wu:2019eyi}
H.~Z.~Wu, L.~G.~Pang, X.~G.~Huang and Q.~Wang,
Phys. Rev. Research. \textbf{1}, 033058 (2019)
[arXiv:1906.09385 [nucl-th]].

\bibitem{Wang:2017jpl}
Q.~Wang,
Nucl. Phys. A \textbf{967}, 225-232 (2017)
[arXiv:1704.04022 [nucl-th]].

\bibitem{Liu:2020ymh}
Y.~C.~Liu and X.~G.~Huang,
Nucl. Sci. Tech. \textbf{31}, 56 (2020)
[arXiv:2003.12482 [nucl-th]].

\bibitem{Huang:2020xyr}
X.~G.~Huang,
[arXiv:2002.07549 [nucl-th]].

\bibitem{Becattini:2020ngo}
F.~Becattini and M.~A.~Lisa,
[arXiv:2003.03640 [nucl-ex]].

\bibitem{Gao:2020vbh}
J.~H.~Gao, G.~L.~Ma, S.~Pu and Q.~Wang,
Nucl. Sci. Tech. \textbf{31}, 90 (2020)
[arXiv:2005.10432 [hep-ph]].

\bibitem{Muller:2018ibh}
B.~Muller and A.~Schaefer,
Phys. Rev. D \textbf{98}, 071902 (2018)
[arXiv:1806.10907 [hep-ph]].

\bibitem{Guo:2019joy}
Y.~Guo, S.~Shi, S.~Feng and J.~Liao,
Phys. Lett. B \textbf{798}, 134929 (2019)
[arXiv:1905.12613 [nucl-th]].

\bibitem{Guo:2019mgh}
X.~Guo, J.~Liao and E.~Wang,
Sci. Rep. \textbf{10}, 2196 (2020)
[arXiv:1904.04704 [hep-ph]].


\end{thebibliography}
\end{document}